\documentclass[12pt]{article}
\let\ssection=\section
\renewcommand{\section}{\setcounter{equation}{0}\ssection}

\newcommand\mathC{\mkern1mu\raise2.2pt\hbox{$\scriptscriptstyle|$}
        {\mkern-7mu\rm C}} 
\newcommand{\mathR}{{\rm I\! R}}         

\newtheorem{definition}{Definition}[section]
\newtheorem{theorem}{Theorem}[section]

\newcommand\mapup[1]{\Big\uparrow
                        \rlap{$\vcenter{\hbox{$\scriptstyle#1$}}$}}
\newcommand\mapright[1]{\smash{
        \mathop{\mbox{\large{$\longrightarrow$}}}\limits^{#1}}}
\newcommand\bundle[3]{\begin{array}[t]{c}
        {#3}\\ \mapup{#2}\\ {#1}\end{array}}
\newcommand\bundlemap[2]{\begin{array}[t]{c}
\mapright{#2}\\
\phantom{\mapup{}}\\\mapright{#1}\\\end{array}}

\newcommand{\spec}[1]{\ensuremath{\sigma ( \hat #1)}}
\newcommand{\SP}{\ensuremath{{\bf \Sigma}}}
\newcommand{\G}{\ensuremath{{\bf G}}}
\newcommand{\V}{\ensuremath{{\cal V}}}

\begin{document}
\begin{titlepage}
\hspace{10truecm}Imperial/TP/00-01/22

\begin{center}
{\large\bf A Topos Perspective on the
Kochen-Specker Theorem:\\[6pt] IV. Interval Valuations}
\end{center}

\vspace{0.8 truecm}

\begin{center}
        J.~Butterfield\footnote{email: jb56@cus.cam.ac.uk;
            jeremy.butterfield@all-souls.oxford.ac.uk}\\[10pt] All
            Souls College\\ Oxford OX1 4AL
\end{center}
\begin{center}
and
\end{center}
\begin{center}
     C.J.~Isham\footnote{email: c.isham@ic.ac.uk}\\[10pt]
        The Blackett Laboratory\\ Imperial College of Science,
        Technology \& Medicine\\ South Kensington\\ London SW7 2BZ\\
\end{center}

\begin{center}
        24 July 2001
\end{center}

\vspace{0.8 truecm}

\begin{abstract}
We extend the topos-theoretic treatment given in previous papers
\cite{IB98,IB99,HIB00} of assigning values to quantities in
quantum theory. In those papers, the main idea was to assign a
sieve as a partial and contextual truth value to a proposition
that the value of a quantity lies in a certain set $\Delta
\subseteq \mathR$. Here we relate such sieve-valued valuations to
valuations that assign to quantities subsets, rather than single
elements, of their spectra (we call these `interval' valuations).
There are two main results. First, there is a natural
correspondence between these two kinds of valuation, which uses
the notion of a state's support for a quantity (Section 3).
Second, if one starts with a more general notion of interval
valuation, one sees that our interval valuations based on the
notion of support (and correspondingly, our sieve-valued
valuations) are a simple way to secure certain natural properties
of valuations, such as monotonicity (Section 4).
\end{abstract}
\end{titlepage}

\section{Introduction}
\label{Sec:Introduction} In three previous papers \cite{IB98,
IB99, HIB00} we have developed a topos-theoretic perspective on
the assignment of values to quantities in quantum theory. In
particular, it was shown that the Kochen-Specker theorem
\cite{KS67} (which states the impossibility of assigning to each
bounded self-adjoint operator on a Hilbert space of dimension
greater than $2$, a real number such that functional relations
are preserved) is equivalent to the non-existence of any global
elements of a certain presheaf \SP, called the `spectral
presheaf'. This presheaf is defined in closely analogous ways on
the category $\cal O$ of bounded self-adjoint operators on a
Hilbert space $\cal H$ (cf.\ \cite{IB98, IB99}), and on the
category $\cal V$ of commutative von Neumann subalgebras of the
algebra of bounded operators on $\cal H$ (cf.\
\cite{HIB00}).\footnote{There is also a closely analogous
presheaf---the dual presheaf $\bf D$---that is defined on the
category $\cal W$ of Boolean subalgebras of the lattice ${\cal
L(H)}$ of projectors on $\cal H$; and the Kochen-Specker theorem
is also equivalent to the non-existence of any global elements of
$\bf D$ (cf.\ \cite{IB98, IB99}). But we shall not discuss $\cal
W$ further in this paper.}

A key result of \cite{IB98, IB99, HIB00} is that, notwithstanding
the Kochen-Specker theorem, it {\em is\/} possible to define
`generalised valuations' on all quantities, in which any
proposition ``$A \in \Delta$'' (read as saying that the value of
the physical quantity $A$ lies in the Borel set of real numbers
$\Delta$) is assigned, in effect, a set of quantities that are
coarse-grainings (functions) of $A$. To be precise, such a
proposition is assigned as a truth value a certain set of
morphisms in the category $\cal O$ (or $\cal V$), this set being
required to have the structure of a {\em sieve\/}. These
generalised valuations can be motivated from various different
perspectives (cf.\ also \cite{B01}). In particular, they obey a
condition analogous to the {\em FUNC\/} condition of the
Kochen-Specker theorem, which states that assigned values preserve
functional relations between operators, and certain other natural
conditions too. Furthermore, each (pure or mixed) quantum state
defines such a valuation. In Section \ref{Sec:Von} we will briefly
recall the details of these proposals and results.

In this paper, we shall extend this treatment in two main ways
(Sections \ref{Sec:interval} and \ref{Sec:ValSubobjects}
respectively). Both involve the relation between sieve-valued
valuations and valuations that assign to a quantity $A$, not an
individual member of its spectrum, but rather some subset of it;
(which we call `interval valuations'). Though this idea seems at
first sight very different from our generalised valuations---that
assign sieves to propositions ``$A \in \Delta$''---the two types
of valuations turn out to be closely related. In fact, there is a
natural correspondence between them that uses the notion of the
{\em support\/} of a state for a quantity (Section
\ref{Sec:interval}). This correspondence is best expressed for
the case of $\cal V$ than for $\cal O$ since, by using von
Neumann algebras as the base category, various measure-theoretic
technicalities can be immediately dealt with. However, we shall
also discuss the case of $\cal O$, as it is heuristically
valuable.

In Section \ref{Sec:ValSubobjects}, we describe how if one starts
with a yet more general notion of an interval valuation ({\em
i.e.}, one that does not appeal to the notion of support), one
sees that our interval valuations based on the notion of support
(and correspondingly, our sieve-valued valuations) are a simple
way to secure certain natural properties of valuations, such as
monotonicity.

\section{Review of Our Framework}
\label{Sec:Von}
\subsection{The categories $\cal O$ and $\cal V$}
We will first summarise the definitions given in the previous
papers \cite{IB98,IB99,HIB00}, of the categories $\cal O$ and
$\cal V$ that are defined in terms of the operators on a Hilbert
space, and over which various presheaves may be usefully
constructed.

The objects of the category $\cal O$ are defined to be the
bounded self-adjoint operators on the Hilbert space $\cal H$ of
some quantum system. A morphism $f_{{\cal O}}: \hat B \rightarrow
\hat A$ is defined to exist if $\hat B = f(\hat A)$ for some
Borel function $f$. On the other hand, the objects $V$ of the
category $\cal V$ are defined to be the commutative von Neumann
subalgebras of the algebra $B({\cal H})$ of bounded operators on
$\cal H$. The morphisms in $\cal V$ are the subset
inclusions---so if $V_2\subseteq V_1$, we have a morphism
$i_{V_2V_1}: V_2 \rightarrow V_1$. Thus the objects in the
category $\cal V$ form a poset.

The category $\cal V$ gives the most satisfactory description of
the ordering structure of operators. Some reasons for this were
discussed in Section 2.1 of \cite{HIB00}. In particular, each von
Neumann algebra contains the spectral projectors of all its
self-adjoint members; so in a sense $\cal V$ subsumes both $\cal
O$ and $\cal W$ (the category of Boolean subalgebras of the
lattice ${\cal L}({\cal H})$). More important in this paper is
the fact that in \V\ issues about measure theory and spectral
theory are easier to treat than they are in $\cal O$; though we
will for heuristic reasons keep $\cal O$ in play.

\subsection{The spectral presheaf on $\cal O$ and $\cal V$}
The spectral presheaf \SP\ over $\cal O$ was introduced in
\cite{IB98}. It assigns to each object $\hat A$ in $\cal O$ its
spectrum $\sigma({\hat A})$ as a self-adjoint operator:  thus
${\bf \Sigma}({\hat A}) : = \sigma({\hat A}) \subseteq \mathR$.
And it assigns to each morphism $f_{{\cal O}}: \hat B \rightarrow
\hat A$ (so that $\hat B = f(\hat A)$), the corresponding map from
$\sigma({\hat A})$ to $\sigma({\hat B})$: ${\bf \Sigma}(f_{{\cal
O}}): \sigma({\hat A}) \rightarrow \sigma({\hat B})$, with
$\lambda \in \sigma({\hat A}) \mapsto f(\lambda) \in \sigma({\hat
B})$.

We now define the corresponding presheaf over $\cal V$. We recall
(see for example, \cite{KR83a}) that the spectrum $\sigma(V)$ of
a commutative von Neumann algebra $V$ is the set of all
multiplicative linear functionals $\kappa : V \rightarrow
\mathC$; these are also the pure states of $V$. Such a functional
assigns a complex number $\kappa(\hat A)$ to each operator $\hat
A \in V$, such that $\kappa(\hat A) \kappa (\hat B) = \kappa
(\hat A \hat B)$. If $\hat A$ is self-adjoint then $\kappa(\hat
A)$ is real and belongs to the spectrum $\sigma({\hat A})$ of
$\hat A$.

Furthermore, $\sigma(V)$ is a compact Hausdorff space when it is
equipped with the weak-$*$ topology, which is defined to be the
weakest topology such that, for all $\hat A\in V$, the map
$\tilde A :\sigma(V)\rightarrow\mathC$ defined by
\begin{equation}
\tilde A(\kappa):= \kappa(\hat A) ,
                \label{Def:GelTransf}
\end{equation}
is continuous. The function $\tilde A$ defined in Eq.\
(\ref{Def:GelTransf}) is known as the {\em Gelfand transform\/}
of $\hat A$, and the spectral theorem for commutative von Neumann
algebras asserts that the map $\hat A\mapsto \tilde A$ is an
isomorphism of $V$ with the algebra $C(\sigma(V))$ of
complex-valued, continuous functions on $\sigma(V)$. The spectral
presheaf is then defined as follows.

\begin{definition}\label{Defn:spectral}
The {\em spectral presheaf} over $\cal V$ is the contravariant
functor ${\bf \Sigma} : {\cal V} \rightarrow {\rm Set}$ defined
as follows:
\begin{itemize}
\item On objects: ${\bf\Sigma}(V):=\sigma(V)$,
where $\sigma(V)$ is the spectrum of the commutative von Neumann
algebra $V$, {\em i.e.} the set of all multiplicative linear
functionals $\kappa:V \rightarrow \mathC$.

\item On morphisms: If $i_{V_2 V_1}:V_2\rightarrow V_1$,
 so that $V_2 \subseteq V_1$, then ${\bf\Sigma}(i_{V_2
V_1}):\sigma(V_1)\rightarrow \sigma(V_2)$ is defined by
${\bf\Sigma}(i_{V_2 V_1}) (\kappa):= \kappa|_{V_2}$, where
$\kappa|_{V_2}$ denotes the restriction of the functional
$\kappa: V_1 \rightarrow \mathC$ to the subalgebra $V_2$ of $V_1$.
\end{itemize}
\end{definition}

    When restricted to the self-adjoint elements of $V$,  a
multiplicative linear functional $\kappa$ satisfies all the
conditions of a {\em valuation}, namely:
\begin{enumerate}
\item the (real) value  $\kappa(\hat A) $ of $\hat A$ must belong to the
spectrum of $\hat A$;
\item the functional composition principle ({\em FUNC})
\begin{equation}
 \kappa(\hat B) = f(\kappa(\hat A)) \label{func}
\end{equation}
holds for any self-adjoint operators $\hat A, \hat B \in V $ such
that $\hat B = f(\hat A)$.
\end{enumerate}
It follows that the Kochen-Specker theorem can be expressed as
the statement that (for $\dim{\cal H}>2$) the presheaf \SP\ over
$\cal V$ has no global elements. Indeed, a global element of \SP\
over $\cal V$ would assign a multiplicative linear functional
$\kappa:V\rightarrow \mathC$ to each commutative von Neumann
algebra $V$ in ${\cal V}$ in such a way that these functionals
`match up' as they are mapped down the presheaf: {\em i.e.}, the
functional $\kappa$ on $V$ would be obtained as the restriction
to $V$ of the functional $\kappa_1 : V_1 \rightarrow \mathC$ for
any $V_1 \supseteq V$. So such a global element would yield a
global valuation, obeying {\em FUNC} Eq.\ (\ref{func}), on all
the self-adjoint operators: a valuation that is forbidden by the
Kochen-Specker theorem.

\subsection{The coarse-graining presheaf on $\cal O$ and $\cal V$}
\label{SubSec:CGPresheaf} Our chief concern is with  {\em
generalised\/} valuations, which are {\em not\/} excluded by the
Kochen-Specker theorem. These are given by first introducing
another presheaf---the {\em coarse-graining\/}---presheaf, which
gives us a structured collection of propositions about the values
of quantities. We summarise the ideas behind the coarse-graining
presheaf in this subsection and the next. Then we use the
topos-theoretic idea of the subobject classifier to assign sieves
as partial, and contextual, truth values to these propositions
(Section \ref{ssec:sievegenval}).

We begin by representing the proposition ``$A \in \Delta$''---that
the value of the physical quantity $A$ lies in the Borel set
$\Delta \subseteq \sigma({\hat A}) \subset \mathR$---by the
corresponding spectral projector for $\hat A$, ${\hat E}[A \in
\Delta]$ (as we will see below, and as discussed in detail in
\cite{HIB00}, Section \ref{Sec:interval}, this last statement
needs to be qualified as regards $\cal V$). The coarse-graining
presheaf is then defined so as to reflect the behaviour of these
propositions as they are mapped between the different stages of
the base category. Specifically, the coarse-graining presheaf over
$\cal O$ is defined (\cite{IB98}, Defn.\ 4.3) as the following
contravariant functor ${\bf G}:{\cal O}\rightarrow{\rm Set}$:
\begin{itemize}
\item {\em On objects in $\cal O$:} ${\bf G}(\hat A):=
W_A$, where $W_A$ is the spectral algebra of $\hat A$ ({\em
i.e.}, the set of all spectral projectors for $\hat A$);

\item {\em On morphisms in $\cal O$:} If $f_{\cal O}:\hat B
\rightarrow\hat A$ ({\em i.e.}, $\hat B=f(\hat A)$), then ${\bf
G}(f_{\cal O}): W_A\rightarrow W_B$ is defined as
\begin{equation}
        {\bf G}(f_{\cal O})(\hat E[A\in\Delta]):=
        \hat E[f(A)\in f(\Delta)]       \label{Def:G(O)} .
\end{equation}
\end{itemize}
Note that the action of this presheaf coarsens propositions (and
their associated projectors) in the sense that, in the
partial-ordering of the lattice of projectors, $\hat E[f(A)\in
f(\Delta)] \ge \hat E[A\in \Delta]$, and where the strict
inequality arises when $f$ is not injective.

One subtlety is that for $\Delta$ a Borel subset of \spec{A},
$f(\Delta)$ need not be Borel. This is resolved in \cite{IB98},
theorem (4.1), by using the fact that if $\hat A$ has a purely
discrete spectrum (so that, in particular, $f(\Delta)$ {\em is\/}
Borel) then
\begin{equation}
\hat E[f(A)\in f(\Delta)] =\inf \{\hat Q\in W_{f(A)}\subseteq W_A
    \mid \hat E[A\in\Delta]\leq \hat Q\}
                        \label{Theorem:inf}
\end{equation}
where the infimum of projectors is taken in the (complete)
lattice structure of all projectors on $\cal H$. This motivated
the use in \cite{IB98} of Eq.\ (\ref{Theorem:inf}) to {\em
define} the coarse-graining operation for a general self-adjoint
operator $\hat A$: {\em i.e.}, the projection operator denoted by
$\hat E[f(A)\in f(\Delta)]$ is {\em defined\/} using the right
hand side of Eq.\ (\ref{Theorem:inf}).

This infimum construction is used again in our definition of $\bf
G$ over $\cal V$. Specifically, we define:
\begin{definition}\label{Defn:coarse}
The {\em coarse-graining presheaf} over $\cal V$ is the
contravariant functor ${\bf G}: {\cal V} \rightarrow {\rm Set}$
defined as follows:
\begin{itemize}
\item On objects: ${\bf G}(V)$ is the lattice ${\cal L}(V)$ of projection
operators in the commutative von Neumann algebra $V$.

\item On morphisms: if $i_{V_2V_1}: V_2 \rightarrow V_1$ then ${\bf
G}(i_{V_2V_1}): {\cal L}(V_1)\rightarrow {\cal L}(V_2)$ is the
coarse-graining operation defined on $\hat P \in {\cal L}(V_1)$ by
\begin{equation}
{\bf G}(i_{V_2V_1})(\hat P):= {\rm inf}\{\hat Q \in {\cal L}(V_2)
\mid \hat P \leq i_{V_2V_1}(\hat Q) \}
\end{equation}
where the infimum exists because $ {\cal L}(V_2) $ is complete.
\end{itemize}
\end{definition}

The spectral and coarse-graining presheaves will play a central
role in the subsequent discussion. First, however, we recall that
the interpretation of the propositions ``$A \in \Delta$'' is more
subtle for the base category $\cal V$ than for $\cal O$ (as
discussed in \cite{IB98,IB99}). For we interpret a projector
$\hat P \in {\cal L}(V)$ as a proposition about the entire stage
$V_1$. Formally, we can make this precise in terms of the
spectrum of the {\em algebra\/} $V$. That is to say, we note that:

\begin{itemize}
\item Any projector $\hat P\in {\cal L}(V)$ corresponds not only to a 
subset of the spectrum of individual operators $\hat A \in V$
(where $\hat P \in W_A$ so $\hat P = \hat E[A \in \Delta]$ for
some $\Delta \subset \spec{A}$), but also to a subset of the
spectrum of the whole algebra $V$, namely, those multiplicative
linear functionals $\kappa : V\rightarrow \mathC$ such that
$\kappa (\hat P ) = 1$. It will be convenient in Section
\ref{ssec:power} {\em et seq.} to have a notation for this
subset, so we define $V({\hat P}) := \{\kappa \in \sigma(V):V
\rightarrow \mathC \mid \kappa({\hat P}) = 1 \}$.

\item Coarse-graining respects this interpretation in the sense that
if we interpret $\hat P \in {\cal L}(V_1)$ as a proposition about
the spectrum of the algebra $V_1$, then the coarse-graining of
$\hat P$ to some $V_2 \subset V_1$, given by ${\rm inf}\{ \hat Q
\in {\cal L}(V_2) \mid \hat P \leq i_{V_2V_1}(\hat Q) \}$ is a
member of ${\cal L}(V_2)$, and so can be interpreted as a
proposition about the spectrum of the algebra $V_2$.
\end{itemize}

This treatment of propositions as concerning the spectra of
commutative von Neumann algebras, rather than the spectra of
individual operators, amounts to the semantic identification of
all propositions in the algebra corresponding to the same
mathematical projector. Thus when we speak of a proposition ``$A
\in \Delta$'' at some stage $V$, with $\hat A\in V$, we really
mean the corresponding proposition about the spectrum of the
whole algebra $V$ defined using the projector $\hat E[A \in
\Delta]$. In terms of operators, the proposition ``$A \in
\Delta$'' is {\em augmented\/}: it can be thought of as the family
of propositions ``$B \in \Delta_B$'' about operators $\hat B \in
V$ such that the projector $\hat E[A \in \Delta]$ belongs to the
spectral algebra of $\hat B$, and $\hat E[A \in \Delta] = \hat
E[B \in \Delta_B]$.

\subsection{{\bf G} as the Power Object of \SP}\label{ssec:power}
In any topos, any object $X$ has an associated `power object' $PX
:= \Omega^X$, which is the topos analogue of the power set of a
set. Accordingly, in a topos of presheaves, any presheaf $\bf X$
has a  `power presheaf' $P({\bf X})$. This presheaf $P({\bf X})$
assigns to each stage $A$ of the base category the power set
$P({\bf X}(A))$ of the set ${\bf X}(A)$ assigned by $\bf X$; and
it assigns to each morphism $f:B \rightarrow A$ in the
base-category, the set-function from $P({\bf X}(A))$ to $P({\bf
X}(B))$ induced in the obvious way from ${\bf X}(f):{\bf
X}(A)\rightarrow {\bf X}(B)$. That is to say, $P({\bf
X})(f):P({\bf X}(A)) \rightarrow P({\bf X}(B))$ is defined by
\begin{equation}
P({\bf X})(f):\Delta \in P({\bf X}(A)) \mapsto {\bf
(X}(f))(\Delta) \in P({\bf X}(B)). \label{eqn;powermorphism}
\end{equation}

But one can also consider more restricted power presheaves, whose
assignment at a stage $A$ contains only certain subsets of the
set ${\bf X}(A)$. Indeed, in \cite{IB98} Section 4.2.3, it was
noted that the coarse-graining presheaf \G\ over $\cal O$ was
essentially the same as the presheaf $B\SP$ over $\cal O$, which
is defined as assigning to each $\hat A$ in $\cal O$ the Borel
subsets of the spectrum of $\hat A$. The presheaf $B\SP$ on $\cal
O$ was essentially the Borel power object of \SP, containing those
subobjects of \SP\ which are formed of Borel sets of spectral
values, with a projector $\hat E[A \in \Delta] \in {\bf G}(\hat
A)$ corresponding to the Borel subset $\Delta \subset \SP(\hat
A)$.

This connection between projectors and subsets of spectra---{\em
i.e.}, the fact that $\bf G$ is essentially the same as a
restricted power presheaf of $\SP$---also holds in the case of
$\cal V$, as follows.

A projection operator $\hat P \in V$ has as its Gelfand transform
$\tilde P$ the characteristic function of a subset of the
spectrum of $V$, namely the set $V({\hat P})$ of multiplicative
linear functionals $\kappa$ on $V$ such that $\kappa(\hat P) =
1$; in other words, ${\tilde P} = \chi_{V({\hat P})}$. This set is
both closed and open ({\em clopen}) in the compact Hausdorff
topology of $\sigma(V)$. Conversely, each clopen subset of
$\sigma(V)$ corresponds to a projection operator $\hat P$ whose
representative function $\tilde P$ on $\sigma(V)$ is the
characteristic function of this subset.

So in analogy with $B\SP$ on $\cal O$, we define a similar
presheaf on $\cal V$, viz.\ the clopen power object of \SP, which
we will denote ${\rm Clo}{\bf\Sigma}$:
\begin{itemize}
\item On objects: ${\rm Clo}\SP(V)$ is defined to be the
set of clopen subsets of the spectrum $\sigma(V)$ of the algebra
$V$. Each such clopen set is the set $V({\hat P})$ of
multiplicative linear functionals $\kappa$ such that $\kappa
(\hat P) = 1$ for some projector $\hat P \in V$. So ${\rm
Clo}\SP(V) = \{V({\hat P})\mid {\hat P} \in {\cal L}(V)\}$.

\item On morphisms: for $V_2 \subset V_1$, we define in accordance with 
Eqn (\ref{eqn;powermorphism})
\begin{eqnarray}
{\rm Clo}\SP(i_{V_2V_1})(V_1({\hat P})):= {\rm
Clo}\SP(i_{V_2V_1})\left( \{ \kappa \in
\sigma(V_1) \mid \kappa (\hat P)= 1\} \right) \nonumber \\
= \{\lambda \in \sigma(V_2) \mid \lambda = \kappa \mid_{V_2}
\mbox{ for some } \kappa \in V_1({\hat P}) \}
\label{eqn:Clomorphism}
\end{eqnarray}
\end{itemize}

The right hand side of Eq.\ (\ref{eqn:Clomorphism}) is clearly a
subset of $\{ \lambda \in \sigma(V_2) \mid \lambda ({\bf
G}(i_{V_2V_1})(\hat P))= 1\}$. But the converse inclusion also
holds: {\em i.e.}, any $\lambda \in \sigma(V_2)$ such that
$\lambda({\bf G}(i_{V_2V_1})(\hat P)) = 1$ is the restriction to
$V_2$ of some $\kappa \in \sigma(V_1)$ such that $\kappa({\hat
P}) = 1$. This follows from Theorem 4.3.13 (page 266) of
\cite{KR83a}.\footnote{We thank Hans Halvorson for this
reference.} This theorem concerns extending states from a
subspace of a {\em C}*-algebra to the whole {\em C}*-algebra. But
using the fact that the set of pure states of a commutative {\em
C}*-algebra is its spectrum, the theorem, especially part (iv),
implies that $\lambda \in \sigma(V_2)$ can be extended to a
$\kappa \in \sigma(V_1)$, with $\kappa$  chosen so that $\kappa
(\hat P) = c$ for any $c$ such that:
\begin{equation}
c \geq \mbox{sup} \{ \lambda(\hat E)\mid \mbox{$\hat E$ is a
projector in $V_2$ and ${\hat E} \leq {\hat P}$}
\}\label{eqn:cgreater}
\end{equation}
\begin{equation}
c \leq \mbox{inf} \{ \lambda(\hat E)\mid \mbox{$\hat E$ is a
projector in $V_2$ and ${\hat P} \leq {\hat E}$}
\}\label{eqn:clesser}
\end{equation}
For our case of $c = 1$, the first constraint Eq.\
(\ref{eqn:cgreater}) is trivial.  And the infimum in the second
constraint, Eq.\ (\ref{eqn:clesser}), is 1; for the fact that
${\bf G}(i_{V_2V_1})(\hat P)) \leq {\hat E}$ for all ${\hat E}
\in V_2$ such that ${\hat P} \leq {\hat E}$ implies that $\lambda
({\hat E})=1$ for all ${\hat E} \in V_2$ such that ${\hat P} \leq
{\hat E}$. So $\lambda$ has an extension $\kappa$ such that
$\kappa(\hat P) = 1$.

This result, that $\{\lambda \in \sigma(V_2) \mid \lambda =
\kappa|_{V_2} \mbox{ for some } \kappa \in V_1({\hat P}) \} = \{
\lambda \in \sigma(V_2) \mid \lambda ({\bf G}(i_{V_2V_1})(\hat
P))= 1\}$, implies that $\bf G$ and ${\rm Clo}\SP$ are isomorphic
in the topos of presheaves on $\cal V$. Here isomorphism means as
usual that: (i) there is a natural transformation $N$ from  $\bf
G$ to ${\rm Clo}\SP$, {\em i.e.}, a family of maps
$N_V:\G(V)\rightarrow{\rm Clo}\SP(V)$ for each stage $V$ in $\cal
V$ such that the diagram for $V_2 \subseteq V_1$:
\begin{equation}
    \bundle{{\bf G}(V_1)}{N_{V_1}}{{\rm Clo}\SP(V_1)}
    \bundlemap{{\bf G}(i_{V_2V_1})}{{\rm restriction}}
    \bundle{{\bf G}(V_2)}{N_{V_2}}{{\rm Clo}\SP(V_2)}    \label{diag}
\end{equation}
commutes; and (ii) $N$ is invertible.

Such a natural transformation is provided by $N_V:{\hat P} \in \G
\mapsto V({\hat P}) \in {\rm Clo}\SP(V)$. With this definition of
$N$, the requirement that the diagram in (\ref{diag}) commutes is
\begin{equation}
V_1({\hat P})|_{V_2} = V_2[\G(i_{V_2V_1})({\hat P})]
\label{eqn:isoaseqn}
\end{equation}
which is just the result that $\{\lambda \in \sigma(V_2) \mid
\lambda = \kappa|_{V_2} \mbox{ for some } \kappa \in V_1({\hat
P}) \} = \{ \lambda \in \sigma(V_2) \mid \lambda ({\bf
G}(i_{V_2V_1})(\hat P))= 1\}$. This natural transformation $N$ is
invertible since any clopen set $\in {\rm Clo}\SP(V)$ is of the
form $V({\hat P})$ for a unique projector ${\hat P} \in \G(V)$.

To sum up, we can think of $\bf G$ on $\cal V$ as being the
clopen power object of \SP\ on $\cal V$. This isomorphism will be
important in Section \ref{Sec:interval}.

\subsection{Sieve-Valued Generalised Valuations} \label{ssec:sievegenval}
We now describe how to use the topos-theoretic idea of a
subobject classifier to assign sieves as partial and contextual
truth values to the propositions provided by the coarse-graining
presheaf; (propositions that, for $\cal V$ as base category, are
``augmented'' in the sense discussed at the end of Section
\ref{SubSec:CGPresheaf}).

As mentioned in Section \ref{Sec:Introduction}, these {\em
sieve}-valued valuations have certain properties which strongly
suggest that they are appropriate generalisations of the usual
idea of a valuation. In particular, they satisfy a functional
composition principle analogous to Eq.\ (\ref{func}).
Furthermore, these valuations can be motivated from various
different perspectives, discussed in \cite{IB98, IB99, B01}. The
main motivation, which applies equally to either $\cal O$ or $\cal
V$ as base category, lies in the facts that:
\begin{enumerate}
\item For any base-category, $\cal C$ say, the subobject classifier ${\bf
\Omega}$ in the topos of presheaves $\rm Set^{{\cal C^{\rm op}}}$
is the presheaf that (i) assigns to each $A$ in $\cal C$ the set
${\bf \Omega}(A)$ of all sieves on $A$, where a sieve on $A$ is a
set of morphisms in $\cal C$ with codomain $A$ that is `closed
under composition', {\em i.e.} if $S$ is sieve on $A$ and
$f:B\rightarrow A$ is a morphism in $S$, then for any
$g:C\rightarrow B$, the composite $f\circ g:C\rightarrow A$ is in
$S$; and (ii) assigns to each morphism $f:B\rightarrow A$ in $\cal
C$, the pullback map on sieves ${\bf \Omega}(f):{\bf
\Omega}(A)\rightarrow {\bf \Omega}(B)$; (cf.\ \cite{IB98},
Appendix).

We note that the base-category of most interest for us, $\cal V$,
is a poset; and since in a poset there is at most one morphism
between objects, sieves can be identified with lower sets in the
poset. Thus on $\cal V$, the subobject classifier ${\bf \Omega}$
is as follows:
\begin{itemize}
\item {On objects:
${\bf\Omega}(V)$ is the set of sieves in $\cal V$ on $V$. We
recall that ${\bf\Omega}(V)$ has (i) a minimal element, the empty
sieve, $0_V = \emptyset$; and (ii) a maximal element, the
principal sieve, true$_V = \downarrow\!_V:=
\{i_{V'V}:V'\rightarrow V \mid V' \subseteq V\}= \{V' \mid V'
\subseteq V\}$. }

\item On morphisms: ${\bf \Omega}(i_{V_2V_1}):{\bf \Omega}(V_1)
\rightarrow {\bf \Omega}(V_2)$ is the pull-back of the sieves in
${\bf \Omega}(V_1)$ along $i_{V_2V_1}$ defined by:
\begin{eqnarray}
{\bf \Omega}(i_{V_2V_1})(S) = i_{V_2V_1}^* (S) &:=& \{
i_{V_3V_2}:V_3 \rightarrow V_2 \mid
i_{V_2V_1} \circ i_{V_3V_2} \in S \}\\
    &=&\{V_3\subset V_2|V_3\in S\}
\end{eqnarray}
\noindent for all sieves $S \in {\bf \Omega}(V_1)$.
\end{itemize}

\item {In any topos of presheaves $\rm Set^{{\cal C^{\rm op}}}$, morphisms
from an arbitrary presheaf $\bf X$ to the subobject classifier
${\bf \Omega}$ generalize the characteristic functions $\chi$ in
set theory that map an arbitrary set $X$ to the two classical
truth values $\{0,1\}$. In particular, just as the characteristic
function $\chi_K:X \rightarrow \{0,1\}$ for a given subset $K
\subseteq X$ encodes the answers to the questions for each $x \in
X$, ``$x \in K$?'', so also the morphism $\chi_{\bf K}:\bf X
\rightarrow{\bf \Omega}$ for a given subobject (sub-presheaf)
$\bf K$ of $\bf X$ encodes the answers to the questions for each
stage $A$ in the base-category, and each $x \in {\bf X}(A)$, ``at
what stage does a `descendant' of $x$ enter $\bf K$?''.

Thus the sieves can be considered as generalized---more precisely,
partial and contextual---truth values.}

\end{enumerate}

The actual definition of a sieve-valued generalised valuation on
$\cal V$ is as follows. (The definition on $\cal O$ can be
obtained {\em mutatis mutandis}.)
\begin{definition}\label{Defn:gen-val-V}
A {\em sieve-valued generalised valuation\/} on the category
$\cal V$ in a quantum theory is a collection of maps $\nu_V:{\cal
L}(V) \rightarrow {\bf\Omega}(V)$, one for each `stage of truth'
$V$ in the category $\cal V$, with the following properties:
\end{definition}
\noindent {\em (i) Functional composition}:
\begin{eqnarray}
\lefteqn{\mbox{\ For any ${\hat P} \in {\cal L(V)}$ and any $V'
\subseteq V$, so that $i_{V'V}:V' \rightarrow V$, we have}}
\hspace{3cm}\nonumber
\\[3pt]
  &&\nu_{V'}({\bf G}(i_{V'V}(\hat P))= i_{V'V}^*(\nu_V(\hat P))
\hspace{2cm} \ \label{FC-gen-V}
\end{eqnarray}
where $i_{V'V}^*$ is the pull-back of the sieves in ${\bf
\Omega}(V)$ along $i_{V'V}$ defined by
\begin{eqnarray}
{\bf \Omega}(i_{V'V})(S) := i_{V'V}^* (S) &:=& \{ i_{V''V'}:V''
\rightarrow V' \mid i_{V'V} \circ i_{V''V'} \in S \}
\end{eqnarray}
\noindent for all sieves $S \in {\bf \Omega}(V)$.

\noindent {\em (ii) Null proposition condition}:
\begin{equation}
                \nu_V(\hat 0)=0_V \label{Null-gen-V}
\end{equation}

\noindent {\em (iii) Monotonicity}:
\begin{equation}
        \mbox{If }\hat P,\hat Q\in {\cal L}(V)\mbox{ with }
        \hat P\leq\hat Q,\mbox{ then }
        \nu_V(\hat P)\leq\nu_V(\hat Q). \label{Mono-gen-V}
\end{equation}

\noindent We may wish to supplement this list with:

\noindent\smallskip {\em (iv) Exclusivity}:
\begin{equation}
        \mbox{If $\hat P,\hat Q\in {\cal L}(V)$ with
$\hat P\hat Q=\hat 0$ and $\nu_V(\hat P)= {\rm true}_V$, then
$\nu_V(\hat Q)< {\rm true}_V$} \label{Excl-gen-V}
\end{equation}
and

\smallskip\noindent
{\em (v) Unit proposition condition}:
\begin{equation}
        \nu_V(\hat 1)=\mbox{true}_V.
        \label{unit-prop-cond-V}
\end{equation}

Note that in writing Eq.\ (\ref{FC-gen-V}), we have employed
Definition \ref{Defn:coarse} to specify the coarse-graining
operation in terms of an infimum of projectors, as motivated by
Theorem 4.1 of \cite{IB98}.

The topos interpretation of these generalised valuations remains
as discussed in Section 4.2 of \cite{IB98} and Section 4 of
\cite{IB99}. Adapting the results and discussion to the category
$\cal V$, we have in particular the result that because of the
$FUNC$ condition, Eq.\ (\ref{FC-gen-V}), the maps
$N_V^{\nu}:{\cal L}(V) \rightarrow {\bf \Omega}(V)$ defined at
each stage $V$ by:
\begin{equation}
    N_V^{\nu}(\hat P) = \nu_V(\hat P)
\end{equation}
define a natural transformation $N^{\nu}$ from $\bf G$ to ${\bf
\Omega}$. Since ${\bf \Omega}$ is the subobject-classifier of the
topos of presheaves, ${\rm Set}^{\cal V^{\rm op}}$, these natural
transformations are in one-to-one correspondence with subobjects
of $\bf G$; so that each generalised valuation defines a
subobject of $\bf G$. We will pursue this topic in more detail in
Section \ref{Sec:interval}.

\subsection{Sieve-Valued Valuations Associated with Quantum States}
 \label{ssec:genvalquantumstate}
We recall (for example, \cite{IB98}, Definition 4.5) that each
quantum state $\rho$ defines a sieve-valued generalised valuation
on $\cal O$ in a natural way by
\begin{eqnarray}
 \nu^\rho(A\in\Delta)&:=&\{f_{\cal O}:\hat B\rightarrow \hat A
\mid {\rm Prob}(B\in f(\Delta);\rho)=1\}\nonumber
\\[2pt]
                &\,=&\{f_{\cal O}:\hat B\rightarrow \hat A
\mid {\rm tr}(\rho\,\hat E[B\in f(\Delta)])=1\}.
\label{Def:nurhoDelta}
\end{eqnarray}
 Thus the generalised valuation associates
to the proposition ``$A\in\Delta$'' at stage $\hat A$ all arrows
in $\cal O$ with codomain $\hat A$ along which the projector
corresponding to the proposition coarse-grains to a projector
which is `true' in the usual sense of having a Born-rule
probability equal to $1$, which in our framework corresponds to
the `totally true' truth value, the principal sieve
$\downarrow_{\hat B}$ at stage $\hat B$. This construction is
easily seen to be a sieve, and satisfies conditions analogous to
Eqs.\ (\ref{FC-gen-V}--\ref{unit-prop-cond-V}) for a generalised
valuation on ${\cal V}$ (\cite{IB98}, Section 4.4).

We also recall that there is a one-parameter family of extensions
of these valuations, defined by relaxing the condition that the
proposition coarse-grains along arrows in the sieve to a `totally
true' projector. That is to say, we can define the sieve
\begin{eqnarray}
 \nu^{\rho,r}(A\in\Delta)&:=&\{f_{\cal O}:\hat B\rightarrow \hat A
\mid {\rm Prob}(B\in f(\Delta);\rho) \ge r \}\nonumber
\\[2pt]
                &\,=&\{f_{\cal O}:\hat B\rightarrow \hat A
\mid {\rm tr}(\rho\,\hat E[B\in f(\Delta)]) \ge r\}
\label{Def:nurho-r-Delta}
\end{eqnarray}
 where the proposition ``$A\in\Delta$'' is only required to
coarse-grain to a projector that is true with some probability
greater than $r$, where $0.5 \leq r \leq 1$.

Furthermore, if one drops the exclusivity condition, one can
allow probabilities less than 0.5, {\em i.e.} $0 < r < 0.5$.

Similarly, each quantum state $\rho$ defines a sieve-valued
generalised valuation on $\cal V$ in a natural way. Recall from
Section \ref{SubSec:CGPresheaf} that we interpret a projector
$\hat P \in {\cal L}(V)$ as an ``augmented'' proposition about the
spectrum of the commutative subalgebra $V$, rather than about the
value of just one operator. Thus we define a sieve-valued
generalised valuation associated with a quantum state $\rho$ as
follows:

\begin{definition}\label{Defn:gen-val-V-rho}
The sieve-valued valuation $ \nu^{\rho}_{V_1}$ of a projector
$\hat P \in V_1$ associated with a quantum state $\rho$ is
defined by:
\begin{equation}
\nu^{\rho}_{V_1} (\hat P) := \{ i_{V_2V_1}: V_2 \rightarrow V_1
\mid \rho \: [{\bf G}(i_{V_2V_1})(\hat P)] = 1 \}.
\label{eqn:nurhoV}
\end{equation}
\end{definition}
This assigns as the truth-value at stage $V_1$ of a projector
$\hat P \in {\cal L}(V_1)$, a sieve on $V_1$ containing
(morphisms to $V_1$ from) all stages $V_2$ at which $\hat P$ is
coarse-grained to a projector which is `totally true' in the
usual sense of having Born-rule probability $1$.

One readily verifies that Eq.\ (\ref{eqn:nurhoV}) defines a
generalised valuation in the sense of Definition
\ref{Defn:gen-val-V}. (The verification is the same, {\em mutatis
mutandis}, as for generalised valuations on $\cal O$, given in
Section 4.4 of \cite{IB98}.)

Again, we can obtain a one-parameter family of such valuations by
introducing a probability $r$:
\begin{equation}
\nu^{\rho ,r}_{V_1} (\hat P) := \{ i_{V_2V_1}: V_2 \rightarrow
V_1 \mid \rho \: [{\bf G}(i_{V_2V_1})(\hat P)] \ge r \}
\label{eqn:nurhoV-r}.
\end{equation}

\subsection{Interval Valuations}
\label{SubSec:IntervalValuations} The sieve-valued generalised
valuations on $\cal O$ and $\cal V$ discussed in Sections
\ref{ssec:sievegenval} and \ref{ssec:genvalquantumstate} (and
their analogues on  $\cal W$, discussed in \cite{IB98, IB99}) are
one way of assigning a generalised truth value to propositions in
a way that is not prevented by the Kochen-Specker theorem. We now
turn to relating these to another notion of `generalised
valuation', which we call  `interval valuations' since the
intuitive idea  is to assign some interval of real numbers to
each operator. Note that here `interval' is used loosely: it
means just some (Borel) subset of $\mathR$, not necessarily a
connected subset; and more generally, it means just some (Borel)
subset of the spectrum one is concerned with (at a given stage of
the base-category).

In Section 4 of \cite{HIB00}, we showed how this intuitive idea
can be developed in various ways, even for a single
base-category. That discussion focussed on $\cal V$, and
described how a sieve-valued generalised valuation on $\cal
V$---in particular one associated with a quantum state---induces
an `interval valuation' in various senses of the phrase. These
various senses differ about whether to take the assigned
intervals at the various stages to define:
\begin{enumerate}
\item[] (i) a subobject of \SP, or
\item[] (ii) a global element of $\bf G$, or
\item[] (iii) a subobject of $\bf G$.
\end{enumerate}
But these different senses of `interval valuation' are similar in
that all are defined in terms of the set of `totally true'
projectors at each stage $V$ of the base-category $\cal V$. Thus
for any sieve-valued valuation $\nu$, we defined the {\em truth
set}
\begin{equation}
T^{\nu}(V):= \{{\hat P} \in {\cal L}(V) \mid \nu_{V}({\hat P)} =
\mbox{true}_V\} \label{eqn:truthsetVnu}
\end{equation}
so that, in particular, for the valuation $\nu^\rho$ associated
with the quantum state $\rho$ we have
\begin{equation}
T^{\rho}(V):= \{{\hat P} \in {\cal L}(V) \mid \rho({\hat P}) =
1\}. \label{eqn:truthsetVnurho}
\end{equation}

We used these truth sets in two ways.  First, we defined interval
valuations that are subobjects of \SP\ by assigning to each stage
$V$, the subset of the spectrum $\sigma(V)$ consisting of all
functionals that `make certain' all members of the truth set
$T^{\nu}(V)$. That is, for any sieve-valued valuation $\nu$, we
assign to $V$ the set:
\begin{equation}
{\bf I}^{\nu}(V) = \{ \kappa \in \sigma(V) \mid \kappa (\hat P) =
1, \; \; \forall \hat P \in T^\nu(V) \} = \bigcap_{{\hat P} \in
T^{\nu}(V)}V({\hat P}). \label{eqn:Inu}
\end{equation}
This assignment gives a subobject of \SP\ (meaning that if $V_2
\subseteq V_1$, then ${\bf I}^{\nu}(V_2) \supseteq {\bf
I}^{\nu}(V_1)\mid_{V_2}$) provided the following condition (eqn
(4.9) of \cite{HIB00})
\begin{equation}
 \mbox{ If } V_2 \subseteq V_1 \mbox{ then inf }T^\nu(V_2) \ge
 \mbox{ inf }T^\nu(V_1)
\label{eqn:infcond}
\end{equation}
is satisfied---which it always is for the valuations $\nu^\rho$
associated with quantum states $\rho$ because for these
valuations $T^{\nu^\rho}(V_2) \subseteq T^{\nu^\rho}(V_1)$. It is
also satisfied for the `probability $r$' quantum valuations
$\nu^{\rho, r}$, {\em i.e.} with truth sets defined using Eq.\
(\ref{eqn:nurhoV-r}).\footnote{Eqn.\ (\ref{eqn:Inu}) shows how
`interval' is here used abstractly: an algebra $V$ is assigned a
subset of its spectrum, {\em i.e.} a set of multiplicative linear
functionals on $V$ which corresponds to a subset of the spectrum
of each operator in the algebra.}

Second, we defined interval valuations that are global elements of
$\bf G$ by taking the infima of these truth sets (using the fact
that ${\cal L}(V)$ is a complete lattice) to define what we
called the {\em support} (of the valuation, or the quantum state)
at each stage $V$:
\begin{equation}
s(\nu,V):= \mbox{inf }T^{\nu}(V) = \mbox{inf}\{{\hat P} \in {\cal
L}(V) \mid \nu_{V}({\hat P)} = \mbox{true}_V\}
\label{eqn:supportVnu}
\end{equation}
so that in particular, for the valuation $\nu^\rho$
\begin{equation}
s(\rho,V):= \mbox{inf }T^{\rho}(V) = \mbox{inf}\{{\hat P} \in
{\cal L}(V) \mid \rho({\hat P}) = 1\}. \label{eqn:supportVnurho}
\end{equation}

An example of an interval valuation that is a global element of
$\bf G$ is given by assigning to each stage $V$, the support at
that stage, $s(\nu,V)$ or $s(\rho,V)$. This assignment gives a
global element of $\bf G$ provided that supports (infima of truth
sets) `match up' under coarse-graining in the usual sense that
\begin{eqnarray}
 \mbox{ If } V_2 \subseteq V_1 \mbox{, then }
{\rm inf} \{ \hat P \in T^\nu(V_2) \} =: s(\nu,V_2) &=& {\bf
G}(i_{V_2V_1}) \left( {\rm inf} \{\hat P \in T^\nu(V_1) \}
\right)\nonumber\\
 &=& {\bf G}(i_{V_2V_1})(s(\nu,V_1)). \label{eqn:supportsmatch}
\end{eqnarray}
This condition is satisfied for the valuations $\nu^\rho$
associated with quantum states $\rho$ (but {\em not\/} for the
`probability $r$' quantum valuations $\nu^{\rho, r}$, {\em i.e.}
with supports $s(\nu^{\rho, r},V)$ defined on analogy with Eq.\
(\ref{eqn:supportVnu}) but using Eq.\ (\ref{eqn:nurhoV-r})).

We note that the notion of an interval valuation that is a global
element of $\bf G$ is stronger than the notion of a subobject of
\SP\ (treated in case (i) above) in the sense that any global
element of \G\ defines a subobject of \SP\, but not vice versa.
Thus any global element $\gamma$ of \G---{\em i.e.} an assignment
$\gamma$ such that if $V_2 \subset V_1$ then $\gamma(V_2) ={\bf
G}(i_{V_2V_1})(\gamma(V_1))$---defines a subobject ${\bf
I}^{\gamma}$ of \SP\ by
\begin{equation}
{\bf I}^{\gamma}(V_1) := V_1(\gamma(V_1)) = \{ \kappa \in
\sigma(V_1) \mid \kappa (\gamma(V_1))= 1 \}
\end{equation}
since $\kappa(\hat P)=1$ implies $\kappa({\bf G}(i_{V_2V_1})(\hat
P))=1$, so that ${\bf I}^\gamma(V_1)|_{V_2}\subseteq {\bf
I}^\gamma(V_2)$. We can also put this in terms of the isomorphism
$N$ in Section \ref{ssec:power} between \G\ and ${\rm Clo}\SP$
whose component maps $N_V: {\hat P}\in {\bf G}(V)\mapsto V({\hat
P}) \in {\rm Clo}\SP$ carry the global element $\gamma$ of \G\
into a global element of ${\rm Clo}\SP$, {\em i.e.} a subobject
of \SP.

\section{The Correspondence between Intervals and Sieves}
\label{Sec:interval}
\subsection{Prospectus}
So much by way of review. In the rest of this paper we shall
report some new results about the relation between sieve-valued
valuations and interval valuations; where both of these notions
will be understood more generally than in Sections
\ref{ssec:sievegenval} to \ref{SubSec:IntervalValuations}.
However, in this Section (though not Section
\ref{Sec:ValSubobjects}) all the interval valuations to be
discussed will be like those in Section
\ref{SubSec:IntervalValuations}, in the sense that they will be
based on the notion of truth sets and associated ideas
(especially the infima of truth sets, {\em i.e.},
supports).\footnote{So as in \cite{HIB00}, we are not concerned
here to appeal to interval valuations to solve the measurement
problem, namely by assigning intervals to some or all quantities
that are `narrow' enough to give definite results to quantum
measurements and yet `wide' enough to avoid the Kochen-Specker
and other `no-go' theorems. For a recent discussion of this
strategy, as it occurs within the modal interpretation, cf.\
\cite{Verm00}.}

In this Section, we will discuss a kind of correspondence between
sieve-valued valuations and interval valuations. So despite the
marked differences between sieve-valued valuations and interval
valuations---for example, in $\cal V$ we see projectors or
propositions {\em versus\/} algebras as arguments, and sieves
{\em versus\/} sets of linear functionals as values---it turns
out that they correspond. Indeed, in a sense they mutually
determine each other. We have already seen in Section
\ref{SubSec:IntervalValuations} how sieve-valued valuations
determine interval valuations, via the idea of truth sets. The
converse determination, of sieve-valued valuations by interval
valuations, is simplest for the case where the interval
valuations are global elements of \G; {\em i.e.} for case (ii) of
Section \ref{SubSec:IntervalValuations}, where we use not just
truth sets, but their infima, {\em supports\/}. We shall present
this in Section \ref{sssec:globalG}.

Then in Section \ref{SubSubSec:SubObjectSigma} we shall discuss
case (i) in Section \ref{SubSec:IntervalValuations}, where the
interval valuations are subobjects of \SP.

In both this Section and the next, our discussion will again
concentrate on $\cal V$ since, as mentioned in the Introduction,
using $\cal V$ avoids measure-theoretic difficulties about the
spectra of operators (and functions of them) which arise in $\cal
O$. But since our results about $\cal V$ are rather abstract, it
will be heuristically helpful to report the corresponding claims
about $\cal O$; {\em i.e.\/} to state what our results imply about
presheaves over $\cal O$ at those stages ({\em i.e.} operators) of
$\cal O$  that do not have these measure-theoretic difficulties.
(These stages will include all operators with a pure discrete
spectrum.) So we report these corresponding claims about $\cal O$
in Section \ref{SubSec:CorrespondenceO}.

\subsection{The Correspondence in ${\cal V}$}
\label{SubSec:CorrespondenceV}
\subsubsection{The case of global elements of \G}\label{sssec:globalG}
The correspondence between sieve-valued valuations and interval
valuations is simplest for the case where the interval valuations
are global elements of \G. In each `direction', there is a
natural and simple sufficient condition for correspondence,
satisfied by `most' of the valuations discussed in Sections
\ref{ssec:sievegenval} to \ref{SubSec:IntervalValuations}. More
precisely: given a sieve-valued valuation $\alpha$, and the
corresponding interval valuation, $s^{\alpha}$ say, that $\alpha$
defines in terms of supports, then there is a simple sufficient
condition ((i) below) for $\alpha$ to equal a valuation naturally
defined by $s^{\alpha}$ which takes sets of morphisms as values.
And in the other direction: given an interval valuation $a$ and
the corresponding valuation, $\alpha^a$ say, that $a$ naturally
defines and which takes sets of morphisms as values,  then $a$
equals the interval valuation defined in terms of the supports of
$\alpha^a$; and there is a simple sufficient condition ((ii)
below) for $\alpha^a$ to be sieve-valued. In fact, condition (ii)
is that supports should form a global element of $\bf G$.

But before stating these results it is illuminating to show that,
taken together, conditions (i) and (ii) are also sufficient to
imply that a valuation is an assignment of sieves, and also obeys
$FUNC$. This claim is made precise in Theorem
\ref{Theorem:Val=Sieves}. It shows that conditions (i) and (ii)
taken together are sufficient for an assignment $\alpha$ to each
stage $V$ in \V\ and each ${\hat P} \in {\bf G}(V) := {\cal
L}(V)$ of a set of morphisms $\alpha_V({\hat P}) \subseteq
\{i_{V'V}: V' \rightarrow V \}$, to satisfy three conditions.
Namely, the conditions: (a) that $\alpha$ is sieve-valued ({\em
i.e.} each $\alpha(V)$ is a sieve on $V$); (b) that $\alpha$
obeys $FUNC$; (c) that $\alpha$ obeys a characterization that
encapsulates the correspondence between sieve-valued valuations
and interval valuations. (This characterization will also lead in
to the discussion in Section \ref{Sec:ValSubobjects}.)

We begin by noting that for any such $\alpha$, {\em i.e.} any
such assignments $\alpha_V({\hat P}) \subseteq \{i_{V'V}: V'
\rightarrow V \}$, we can define truth sets and (since ${\cal
L}(V)$ is complete) supports, just as in Eqs.\
(\ref{eqn:truthsetVnu}) and (\ref{eqn:supportVnu}). So we write
these as $T^{\alpha}(V)$ and $s(\alpha,V)$ respectively.
Similarly, for any such $\alpha$, the condition that supports
`match up' under coarse-graining, makes sense;  cf.\ Eq.\
(\ref{eqn:supportsmatch}), substituting $\alpha$ for $\nu$.

\begin{theorem}\label{Theorem:Val=Sieves}
Let $\alpha$ be an assignment to each stage $V$ in \V\ and each
${\hat P} \in {\bf G}(V) := {\cal L}(V)$ of a set of morphisms
$\alpha_V({\hat P}) \subseteq \{i_{V'V}:V'\rightarrow V \}$ with
codomain $V$. Let $T^{\alpha}(V)$ and $s(\alpha,V)$ be defined as
in Eqs.\ (\ref{eqn:truthsetVnu}) and (\ref{eqn:supportVnu})
respectively (just substituting $\alpha$
for $\nu$). Suppose that $\alpha$ obeys:\\[2pt]
(i) If $V_2 \subseteq V_1$ and ${\hat P} \in {\cal L}(V_1)$, then
$i_{V_2V_1}:V_2\rightarrow V_1 \in \alpha_{V_1}({\hat P})$ iff
$s(\alpha,V_2)
\leq {\bf G}(i_{V_2V_1})({\hat P})$;\\[2pt]
(ii) supports give a global element of \G, {\em i.e.} they match
up under coarse-graining, in the sense of Eq.\
(\ref{eqn:supportsmatch}), i.e.\\
\begin{equation}
\mbox{If $V_2 \subseteq V_1$, }s(\alpha,V_2) = {\bf
G}(i_{V_2V_1})(s(\alpha,V_1)). \label{eqn:alphasupportsmatch}
\end{equation}
Then:\\
(a) each $\alpha_V({\hat P})$ is a sieve;\\[2pt]
(b) $\alpha$ obeys $FUNC$, as in Eq.\ (\ref{FC-gen-V}), {\em
i.e.},
\begin{equation}
\alpha_{V_2}({\bf G}(i_{V_2V_1}(\hat P))=
i_{V_2V_1}^*(\alpha_{V_1}(\hat P)) \label{eqn:FC-alpha}
\end{equation}
(c) $\alpha$ obeys
\begin{equation}
\alpha_{V_1}({\hat P}) = \{i_{V_2V_1}:V_2\rightarrow V_1 \mid
{\bf G}(i_{V_2V_1})(s(\alpha,V_1)) \leq {\bf G}(i_{V_2V_1})({\hat
P}) \}. \label{eqn:characterization}
\end{equation}
\end{theorem}

{\bf Proof}: (a): Given $i_{V_2V_1} \in \alpha_{V_1}({\hat P})$,
the condition (i) and the monotonicity of ${\bf G}(i_{V_2V_1})$
imply that for any $V_3 \subseteq V_2$, ${\bf
G}(i_{V_3V_2})(s(\alpha,V_2)) \leq {\bf G}(i_{V_3V_1})({\hat
P})$. But by (ii), ${\bf G}(i_{V_3V_2})(s(\alpha,V_2)) =
s(\alpha, V_3)$; so that by (i), $i_{V_3V_1} \in
\alpha_{V_1}({\hat P})$.

(b): condition (i) implies that $\alpha_{V_2}({\bf
G}(i_{V_2V_1})(\hat P)) = \{i_{V_3V_2}:V_3\rightarrow V_2 \mid
s(\alpha,V_3) \leq {\bf G}(i_{V_3V_1})({\hat P})\}$ and that
$i_{V_2V_1}^*(\alpha_{V_1}(\hat P)) :=
\{i_{V_3V_2}:V_3\rightarrow V_2 \mid i_{V_2V_1} \circ i_{V_3V_2}
\in \alpha_{V_1}({\hat P})\} = \{i_{V_3V_2}:V_3\rightarrow V_2
\mid s(\alpha,V_3) \leq {\bf G}(i_{V_3V_1})({\hat P})\}$. (So
result (b) depends only on the condition (i)).

(c): Immediate: apply (ii) {\em i.e.}, Eq.\
(\ref{eqn:alphasupportsmatch}) to the condition in (i) that
$s(\alpha,V_2) \leq {\bf G}(i_{V_2V_1})({\hat P})$. \hfill QED.

In particular, the sieve-valued valuations associated with
quantum states (for probability 1, but not $r$ with $0 \leq r <1$)
obey the conditions of Theorem \ref{Theorem:Val=Sieves}. For we
noted in Section 4.3 of \cite{HIB00} that (ii) {\em i.e.} Eq.\
(\ref{eqn:supportsmatch}), holds for these valuations; and (i)
holds trivially for them, since $\rho({\bf G}(i_{V_2V_1})({\hat
P})) = 1$ iff $s(\rho,V_2) \leq {\bf G}(i_{V_2V_1})({\hat P})$.

We turn to describing how conditions (i) and (ii) are,
respectively, natural sufficient conditions for: (a) a
sieve-valued valuation to be determined by an interval-valued
valuation that it itself determines; and (b) an interval
valuation to be determined by a sieve-valued valuation that it
itself determines.

First, suppose $\alpha$ is an assignment to each stage $V$ in \V\
and each ${\hat P} \in {\bf G}(V) := {\cal L}(V)$ of a sieve on
$V$. We can define truth sets and (since ${\cal L}(V)$ is
complete) supports,  as in Eqs.\ (\ref{eqn:truthsetVnu}) and
(\ref{eqn:supportVnu}). Let us write these as $T^{\alpha}(V)$ and
$s^{\alpha}(V)$ respectively. Then we define a  valuation with
sets of morphisms as values, in terms of the $s^{\alpha}(V)$, by
\begin{equation}
\alpha^{s^{\alpha}}_{V_1}({\hat P}):= \{i_{V_2V_1}: V_2
\rightarrow V_1 \mid s^{\alpha}(V_2) \leq {\bf
G}(i_{V_2V_1})({\hat P})\}
\end{equation}
and ask: does $\alpha^{s^{\alpha}}_{V_1}({\hat P}) =
\alpha_{V_1}({\hat P})$? The answer is trivially: `Yes' if and
only if $\alpha$ obeys condition (i) of Theorem
\ref{Theorem:Val=Sieves}. (We note incidentally that this
argument, including its definition of truth sets and supports,
does not require that the given $\alpha$ assign sieves. It is
enough, as in Theorem \ref{Theorem:Val=Sieves}, that $\alpha$ be
an assignment to each stage $V$ in \V\ and each ${\hat P} \in
{\bf G}(V) := {\cal L}(V)$ of a set of morphisms with codomain
$V$: the conclusion, that $\alpha^{s^{\alpha}} = \alpha$ iff
$\alpha$ obeys condition (i), is unaffected.)

Second, suppose $a$ is an assignment at each stage $V$ in \V\ of
an element of ${\bf G}(V) := {\cal L}(V)$. (We do not for the
moment require that $a$ define a global element of $\bf G$.) Then
we define a valuation $\alpha^a$, with sets of morphisms as
values, on all $\hat P$ in each ${\bf G}(V) := {\cal L}(V)$, in
terms of $a$, as follows:
\begin{equation}
\alpha^{a}_{V_1}({\hat P}):= \{i_{V_2V_1}: V_2 \rightarrow V_1
\mid a(V_2) \leq {\bf G}(i_{V_2V_1})({\hat P})\}.
\label{eqn:intervaltosieve}
\end{equation}
It follows that the support $s(\alpha^a,V)$, defined in the usual
way (cf.\ Eq.\ (\ref{eqn:supportVnu})) is equal to $a(V)$. That
is, suppose we define truth sets and supports for $\alpha^a$ in
the usual way; cf.\ Eqs.\ (\ref{eqn:truthsetVnu}) and
(\ref{eqn:supportVnu}). Then note that
\begin{equation}
{\hat P} \in T^{\alpha^a}(V)\mbox{ iff } a(V) \leq
\G(i_{VV})({\hat P})= {\hat P},
\end{equation}
so that $s(\alpha^a,V):= \mbox{inf }T^{\alpha^a}(V) = a(V)$.

But under what conditions is $\alpha^a$ sieve-valued ({\em i.e.}
$\alpha^{a}_{V_1}({\hat P})$ is always a sieve)? In fact, the
condition (ii) in Theorem \ref{Theorem:Val=Sieves}---{\em i.e.}
the condition that $a$ defines a global element of $\bf G$---is a
natural sufficient condition for this. For suppose that
$i_{V_2V_1} \in \alpha^a_{V_1}({\hat P})$, {\em i.e.} $a(V_2)
\leq {\bf G}(i_{V_2V_1})({\hat P})$, and pick any
$i_{V_3V_2}:V_3\rightarrow V_2$. Since ${\bf G}(i_{V_3V_2})$ is
monotonic, we get ${\bf G}(i_{V_3V_2})a(V_2) \leq  {\bf
G}(i_{V_3V_1})({\hat P})$. Assuming (ii), {\em i.e.} ${\bf
G}(i_{V_3V_2})a(V_2) = a(V_3)$, it follows that $i_{V_3V_1} \in
\alpha^a_{V_1}({\hat P})$, {\em i.e.} $\alpha^{a}_{V_1}({\hat
P})$ is a sieve.

\subsubsection{The case of subobjects of ${\bf \Sigma}$}
\label{SubSubSec:SubObjectSigma} We return to case (i) of Section
\ref{SubSec:IntervalValuations}.  We recall that for any
sieve-valued valuation $\nu$, the interval valuation that assigns
to each stage $V$ the subset of the spectrum $\sigma(V)$
consisting of all functionals that `make certain' all members of
the truth set $T^{\nu}(V)$, {\em i.e.} the interval valuation of
Eq.\ (\ref{eqn:Inu}):
\begin{equation}
{\bf I}^{\nu}(V) := \{ \kappa \in \sigma(V) \mid \kappa (\hat P)
= 1, \; \; \forall \hat P \in T^\nu(V) \} = \bigcap_{{\hat P} \in
T^{\nu}(V)}V({\hat P})
\end{equation}
defines a subobject of \SP\ provided Eq.\ (\ref{eqn:infcond}) is
satisfied:
\begin{equation}
 \mbox{ If } V_2 \subseteq V_1 \mbox{ then inf }T^\nu(V_2) \ge
 \mbox{ inf }T^\nu(V_1)\mbox{, {\em i.e.} } s(\nu,V_2) \ge
 s(\nu,V_1).
\end{equation}
(Incidentally, this argument does not require that $\nu$ be a
sieve-valued valuation in the strong sense of Definition
\ref{Defn:gen-val-V} (Section \ref{ssec:sievegenval}); it works
for any assignment, to each stage $V$ in \V\ and each ${\hat P}
\in {\bf G}(V) := {\cal L}(V)$, of a sieve on $V$.)

To obtain the analogue of Theorem \ref{Theorem:Val=Sieves} for the
case of subobjects of ${\bf \Sigma}$,  we note that condition (i)
of Theorem \ref{Theorem:Val=Sieves} says that
$i_{V_2V_1}:V_2\rightarrow V_1 \in \alpha_{V_1}({\hat P})$ if and
only if $s(\alpha,V_2) \leq {\bf G}(i_{V_2V_1})({\hat P})$, {\em
i.e.} iff ${\bf G}(i_{V_2V_1})({\hat P})$ is certain at $V_2$
according to $\alpha$. So we expect the corresponding condition,
for a subobject ${\bf I}^\alpha$ of ${\bf \Sigma}$, to be that
${\bf I}^\alpha(V_2) \subseteq V_2({\bf G}(i_{V_2V_1})({\hat P}))$
($\equiv V_1({\hat P})|_{V_2}$ by the isomorphism Eq.\
(\ref{eqn:isoaseqn}) in Section \ref{ssec:power}). Indeed we have:

\begin{theorem}
\label{Theorem:alphaSubobjects} Let $\alpha$ be an assignment to
each stage $V$ in \V\ and each ${\hat P} \in {\bf G}(V) := {\cal
L}(V)$ of a set of morphisms $\alpha_V({\hat P}) \subseteq
\{i_{V'V}:V'\rightarrow V \}$ with codomain $V$. Let
$T^{\alpha}(V)$ and ${\bf I}^{\alpha}(V)$ be defined as in Eqs.\
(\ref{eqn:truthsetVnu}) and (\ref{eqn:Inu}) respectively (just
substituting $\alpha$ for $\nu$). Suppose that $\alpha$
obeys:\\[2pt]
(i) If $V_2 \subseteq V_1$ and ${\hat P} \in {\cal L}(V_1)$, then
$i_{V_2V_1}:V_2\rightarrow V_1 \in \alpha_{V_1}({\hat P})$ iff
${\bf I}^{\alpha}(V_2) \subseteq V_1({\hat
P})\mid_{V_2}$;\\[2pt]
(ii) the intervals ${\bf I}^{\alpha}$ give a `tight' subobject of
\SP\ in the sense that they match up exactly under restriction,
{\em i.e.}
\begin{equation}
\mbox{If $V_2 \subseteq V_1$, then }{\bf I}^{\alpha}(V_2) = {\bf
I}^{\alpha}(V_1)\mid_{V_2}\mbox{: not merely }{\bf
I}^{\alpha}(V_2) \supseteq {\bf I}^{\alpha}(V_1)\mid_{V_2};
\label{eqn:intervaltight}
\end{equation}
Then:\\
(a) each $\alpha_V({\hat P})$ is a sieve;\\[3pt]
(b) $\alpha$ obeys $FUNC$, just as in Eqs.\ (\ref{FC-gen-V}) and
(\ref{eqn:FC-alpha}),  {\em i.e.}
\begin{equation}
\alpha_{V_2}({\bf G}(i_{V_2V_1}(\hat P))=
i_{V_2V_1}^*(\alpha_{V_1}(\hat P)) \label{eqn:FC-alpha-interval}
\end{equation}
(c) $\alpha$ obeys
\begin{equation}
\alpha_{V_1}({\hat P}) = \{i_{V_2V_1}:V_2\rightarrow V_1 \mid
{\bf I}^{\alpha}(V_1)\mid_{V_2} \subseteq V_1({\hat P})\mid_{V_2}
\} \label{eqn:characterization-interval}
\end{equation}
\end{theorem}

\noindent {\bf Proof}: (a): Given $i_{V_2V_1} \in
\alpha_{V_1}({\hat P})$, the condition (i) and the monotonicity
of taking restrictions ({\em i.e.} if $X$ and $Y$ are sets of
functions on a common domain of which $Z$ is a subset, then
$X\subseteq Y$ implies $X|_Z \subseteq Y|_Z$) imply that for any
$V_3 \subseteq V_2$, ${\bf I}^{\alpha}(V_2)|_{V_3} \subseteq
V_1({\hat P})|_{V_3}$. But (ii) implies ${\bf I}^{\alpha}(V_3)
\subseteq {\bf I}^{\alpha}(V_2)|_{V_3}$; so that ${\bf
I}^{\alpha}(V_3) \subseteq V_1({\hat P})|_{V_3}$ and by (i),
$i_{V_3V_1} \in \alpha_{V_1}({\hat P})$.

(b): condition (i) implies that $\alpha_{V_2}({\bf
G}(i_{V_2V_1})(\hat P)) = \{i_{V_3V_2}:V_3\rightarrow V_2 \mid
{\bf I}^{\alpha}(V_3) \subseteq V_2({\bf G}(i_{V_2V_1})({\hat
P}))|_{V_3} \}$. But the isomorphism of $\bf G$ and ${\rm
Clo}\SP$ (cf.\ diagram \ref{diag} and Eq.\ (\ref{eqn:isoaseqn}))
means that $V_2({\bf G}(i_{V_2V_1})({\hat P})) = V_1({\hat
P})|_{V_2}$; restricting both sides of this equation to any $V_3
\subseteq V_2$, we get $V_2({\bf G}(i_{V_2V_1})({\hat P}))|_{V_3}
= V_1({\hat P})|_{V_3}$. On the other hand, condition (i) also
implies that $i_{V_2V_1}^*(\alpha_{V_1}(\hat P)) :=
\{i_{V_3V_2}:V_3\rightarrow V_2 \mid i_{V_2V_1} \circ i_{V_3V_2}
\in \alpha_{V_1}({\hat P})\} = \{i_{V_3V_2}:V_3\rightarrow V_2
\mid {\bf I}^{\alpha}(V_3) \subseteq V_1({\hat P})|_{V_3} \}$.
(So result (b) depends only on the condition (i)).

(c): Immediate: apply (ii) {\em i.e.} Eq.\
(\ref{eqn:intervaltight}) to the condition (i) that ${\bf
I}^{\alpha}(V_2) \subseteq V_1({\hat P})|_{V_2}$. \hfill QED.

We remark that the analogy with Theorem \ref{Theorem:Val=Sieves}
is very close, but we could equally well have proven Theorem
\ref{Theorem:alphaSubobjects} first. Indeed, much of Theorem
\ref{Theorem:alphaSubobjects} can be stated and proved without
mention of \G. More precisely, the coarse-graining map ${\bf
G}(i_{V_2V_1})(\cdot)$, taking projectors ${\hat P} \in {\bf
G}(V_1)$ to projectors in ${\bf G}(V_2)$, is for the most part
replaced by the map $V_1(\cdot)|_{V_2}$, taking projectors ${\hat
P} \in {\bf G}(V_1)$ to subsets of ${\bf \Sigma}(V_2)$. In
particular, only part (b) needs to mention \G\, and to make use
of the isomorphism in Section \ref{ssec:power} between \G\ and
${\rm Clo}\SP$.

In particular, the sieve-valued valuations associated with
quantum states (for probability 1, but not $r$ with $0 \leq r <
1$) obey the conditions of Theorem \ref{Theorem:alphaSubobjects}.
For we noted in Section 4.4.1 of \cite{HIB00} that (ii) {\em
i.e.}, Eq.\ (\ref{eqn:intervaltight}), holds for these
valuations. Besides, (i) holds for these valuations because of
the isomorphism between \G\ and ${\rm Clo}\SP$, specifically Eq.\
(\ref{eqn:isoaseqn}), $V_2({\bf G}(i_{V_2V_1})({\hat P})) =
V_1({\hat P})\mid_{V_2}$; as follows. By the definition of
$\nu^{\rho}$ (cf.\ Section \ref{ssec:genvalquantumstate}),
$i_{V_2V_1}:V_2\rightarrow V_1 \in \nu^{\rho}_{V_1}({\hat P})$
iff ${\bf G}(i_{V_2V_1})({\hat P}) \in T^{\rho}(V_2)$. On the
other hand, condition (i) for $\nu^{\rho}$ is that ${\bf
I}^{\rho}(V_2) \subseteq V_1({\hat P})\mid_{V_2} \equiv V_2({\bf
G}(i_{V_2V_1})({\hat P}))$, {\em i.e.} that if $\kappa \in
\sigma(V_2)$ and $\kappa({\hat P}) = 1, \forall {\hat P} \in
T^{\rho}(V_2)$, then $\kappa({\bf G}(i_{V_2V_1})({\hat P})) = 1$;
which is just that ${\bf G}(i_{V_2V_1})({\hat P}) \in
T^{\rho}(V_2)$.

Furthermore, the discussion of the second half of Section
\ref{sssec:globalG} also carries over {\em mutatis mutandis};
(though there is one difference). That is: conditions (i) and
(ii) are again, respectively, natural sufficient conditions for:
(a) a sieve-valued valuation to be determined by an
interval-valued valuation that it itself determines; and (b) an
interval valuation to be determined by a sieve-valued valuation
that it itself determines.

First, suppose $\alpha$ is an assignment to each stage $V$ in \V\
and each ${\hat P} \in {\bf G}(V) := {\cal L}(V)$ of a sieve on
$V$. We can define truth sets  as in Eq.\
(\ref{eqn:truthsetVnu}), {\em i.e.},
\begin{equation}
T^{\alpha}(V):= \{{\hat P} \in {\cal L}(V) \mid \alpha_V({\hat
P}) = {\rm true}_V \}\label{eqn:Talpha}
\end{equation}
and intervals as in Eq.\ (\ref{eqn:Inu}), {\em i.e.}
\begin{equation}
{\bf I}^{\alpha}(V):= \{\kappa \in \sigma(V) \mid \kappa({\hat
P}) = 1, \forall {\hat P} \in T^{\alpha}(V) \}\label{eqn:Ialpha}
\end{equation}
Then we define a  valuation with sets of morphisms as values, in
terms of the ${\bf I}^{\alpha}(V)$, by:
\begin{equation}
\alpha^{{\bf I}^{\alpha}}_{V_1}({\hat P}):= \{i_{V_2V_1}: V_2
\rightarrow V_1 \mid {\bf I}^{\alpha}(V_2) \subseteq V_1({\hat
P})\mid_{V_2} \};
\end{equation}
and ask: does $\alpha^{{\bf I}^{\alpha}}_{V_1}({\hat P}) =
\alpha_{V_1}({\hat P})$? The answer is trivially: `Yes' if and
only if $\alpha$ obeys condition (i) of Theorem
\ref{Theorem:alphaSubobjects}. (As in Section
\ref{sssec:globalG}, we note incidentally that this argument, including 
its definition of truth sets and intervals, does not require that
the given $\alpha$ assign sieves. It is enough, as in Theorem
\ref{Theorem:alphaSubobjects}, that $\alpha$ be an assignment to
each stage $V$ in \V\ and each ${\hat P} \in {\bf G}(V) := {\cal
L}(V)$ of a set of morphisms with codomain $V$: the conclusion,
that $\alpha^{{\bf I}^{\alpha}} = \alpha$ if and only if $\alpha$
obeys condition (i), is unaffected.)

Second, suppose $a$ is an assignment at each stage $V$ in \V\ of
a subset $a(V)$ of ${\bf \Sigma}(V) := \sigma(V)$. (We do not for
the moment require that $a$ define a subobject of ${\bf
\Sigma}$.) Then we define a valuation $\alpha^a$, with sets of
morphisms as values, on all $\hat P$ in each ${\bf G}(V) := {\cal
L}(V)$, in terms of $a$, as follows:
\begin{equation}
\alpha^{a}_{V_1}({\hat P}):= \{i_{V_2V_1}: V_2 \rightarrow V_1
\mid a(V_2) \subseteq V_1({\hat P})|_{V_2} \}.
\end{equation}
As in Section \ref{sssec:globalG} (after Eq.\
(\ref{eqn:intervaltosieve})), we ask whether the interval ${\bf
I}^{\alpha^a}(V)$, defined in the usual way (cf.\ Eqs.\
(\ref{eqn:Inu}), (\ref{eqn:Talpha}) and (\ref{eqn:Ialpha})) is
equal to $a(V)$. But in Section \ref{sssec:globalG}, the answer
was automatically `Yes'; now it is not. For defining truth sets
and intervals in this way, we get:
\begin{equation}
{\hat P} \in T^{\alpha^a}(V)\mbox{ iff } a(V) \subseteq V({\hat
P})|_V = V({\hat P}),
\end{equation}
so that $\kappa \in \sigma(V)$ is in ${\bf I}^{\alpha^a}(V)$ if
and only if for all $\hat P$ with $a(V) \subseteq V({\hat P})$,
we have $\kappa({\hat P}) = 1$. All elements of $a(V)$ fulfill
this condition so that $a(V) \subseteq {\bf I}^{\alpha^a}(V)$.
But the converse inclusion requires that if $\kappa \notin a(V)$
then there is $\hat Q$ with $a(V) \subseteq V({\hat Q})$ and
$\kappa({\hat Q}) \neq 1$. And in general this will not hold: if
$\kappa$ is in the closure of $a(V)$, but $\kappa \notin a(V)$,
then any clopen (so closed) superset $Y$ of $a(V)$ must contain
$\kappa$; and any such $Y$ is $V({\hat Q})$ for some $\hat Q$. To
get this converse, and so $a(V) = {\bf I}^{\alpha^a}(V)$, the
natural sufficient condition is that the given sets $a(V)$ should
be clopen. (Recall from Section \ref{ssec:power} that every clopen
subset of $\sigma(V)$ corresponds to a projector whose Gelfand
transform on $\sigma(V)$ is the characteristic function of the
subset.)

But under what conditions is $\alpha^a$ sieve-valued ({\em i.e.}
$\alpha^{a}_{V_1}({\hat P})$ is always a sieve)? In fact, the
condition (ii) in Theorem \ref{Theorem:alphaSubobjects}---{\em
i.e.} the condition that $a$ defines a `tight' subobject of ${\bf
\Sigma}$---is a natural sufficient condition for this. For
suppose that $i_{V_2V_1} \in \alpha^a_{V_1}({\hat P})$, so that
$a(V_2) \subseteq V_1({\hat P})|_{V_2}$, and pick any
$i_{V_3V_2}:V_3\rightarrow V_2$. Since restriction is monotonic,
$a(V_2)|_{V_3} \subseteq V_1({\hat P})|_{V_3}$. Assuming (ii),
{\em i.e.} $a(V_2)|_{V_3} = a(V_3)$, it follows that $i_{V_3V_1}
\in \alpha^a_{V_1}({\hat P})$, thus $\alpha^{a}_{V_1}({\hat P})$
is a sieve.

\subsection{The Correspondence in ${\cal O}$}
\label{SubSec:CorrespondenceO} The discussion in Section
\ref{SubSec:CorrespondenceV} is quite abstract. So it is
illuminating to present the same ideas in a more concrete
setting: namely, the valuations on $\cal O$ associated with a
quantum state $\psi \in {\cal H}$, or more generally a density
matrix $\rho$, which (cf.\ Section \ref{ssec:genvalquantumstate},
especially Eq.\ (\ref{Def:nurhoDelta})) are defined by
\begin{equation}
\nu^{\psi}(A \in \Delta) := \{f_{{\cal O}}:{\hat B}\rightarrow
{\hat A}\mid {\hat E}[B \in f(\Delta)]\psi = \psi \}
\label{eqn:nupsionO}
\end{equation}
and
\begin{equation}
\nu^{\rho}(A \in \Delta) := \{f_{{\cal O}}:{\hat B}\rightarrow
{\hat A}\mid \mbox{tr}(\rho{\hat E}[B \in f(\Delta)]) = 1 \}.
\label{eqn:nurhoonO}
\end{equation}

However, as mentioned in Section \ref{Sec:interval}, various
measure-theoretic difficulties about the spectra of operators
(and functions of them) arise in $\cal O$. These centre around
the fact that if ${\hat B} = f({\hat A})$ (so that there is a
morphism $f_{\cal O}:{\hat B}\rightarrow {\hat A}$ in $\cal O$)
then in general, the corresponding spectra (now consisting of
elements of $\mathR$, not of linear functionals on operators!)
have only a subset inclusion
\begin{equation}
f(\sigma({\hat A})) \subseteq \sigma(f({\hat A}))
\label{eqn:Ospectraonlysubset}
\end{equation}
not necessarily an equality; though of course, if $\hat A$ has a
pure discrete spectrum, then $f(\sigma({\hat A})) = \sigma(f({\hat
A}))$.

This situation prompts three further remarks:
\begin{enumerate}
\item For the role of Eq.\ (\ref{eqn:Ospectraonlysubset}) in
defining the spectral presheaf on $\cal O$, cf.\ Section 2 of
\cite{IB98}.

\item As noted in Section 2.1 of \cite{IB98}, the set of
self-adjoint operators on $\cal H$ that have a pure discrete
spectrum is closed under taking functions of its members, and so
forms a base-category ${\cal O}_d$ on which we can define a
spectral presheaf and a coarse-graining presheaf in a manner
exactly parallel to the definitions over $\cal O$.

\item For a more precise statement of the relation of
$f(\sigma({\hat A}))$ and $\sigma(f({\hat A}))$, cf.\ Eq.\ (2.9)
of \cite{IB98}.
\end{enumerate}

To sum up: it will be heuristically helpful to report what the
results in Section \ref{SubSec:CorrespondenceV} imply about
presheaves over $\cal O$ for those operators for which these
measure-theoretic difficulties do {\em not\/} arise. As just
mentioned, this will include all operators with a pure discrete
spectrum; and the rest of this Section can be read as strictly
true for the spectral presheaf and coarse-graining presheaf
defined on ${\cal O}_d$.

We will do this in two stages, in the next two subsections. Both
depend on the following `definition' of what we will call the
{\em elementary support} of a quantum state, relative to a stage
({\em i.e.}, relative to an operator $\hat A$ in $\cal O$). We
say `definition' since the infimum of a family of Borel sets is
not in general Borel, so that the definition applies only in
special cases, in particular in ${\cal O}_d$. (And we say
`elementary', since these supports are, as usual, subsets of
$\mathR$, and we want to emphasise the distinction from the
rigorously defined supports discussed in Section
\ref{SubSec:CorrespondenceV}.)

\begin{definition}\label{Def:ElementarySupport}
The {\em elementary support}, $s(\psi,{\hat A})$, of a vector
state $\psi \in {\cal H}$ for a quantity $\hat A$, is the
smallest set (measure-theoretic niceties apart!) of real numbers
for which $\psi$ prescribes probability 1 of getting a result in
the set, on measurement of the physical quantity $A$. And
similarly for a density matrix $\rho$. More precisely:
\begin{eqnarray}
    s(\psi,{\hat A}) := {\rm inf_{Borel}}\{\Delta \subseteq \mathR \mid
{\hat
E}[A \in \Delta]\psi = \psi \}.\label{Def:support}  \\[2pt]
    s(\rho,{\hat A}) := {\rm inf_{Borel}}\{\Delta \subseteq \mathR \mid 
{\rm tr}[\rho{\hat E}[A \in \Delta]] = 1 \}.\nonumber
\end{eqnarray}
\end{definition}

\subsubsection{Characterizing quantum valuations with elementary supports}
Given this definition of supports, we can deduce a
characterization of the sieve-valued valuations on $\cal O$
associated with quantum states as defined in Eqs.\
(\ref{eqn:nupsionO}) and (\ref{eqn:nurhoonO}). This
characterization is the analogue of those in Theorem
\ref{Theorem:Val=Sieves}, {\em i.e.}, Eq.\
(\ref{eqn:characterization}) of part (c), and in Theorem
\ref{Theorem:alphaSubobjects}, {\em i.e.}, Eq.\
(\ref{eqn:characterization-interval}); and this characterization,
being more concrete, is heuristically valuable.

For any ${\hat A}$, $\Delta,$ and $f:\Delta \subseteq
f^{-1}(f(\Delta))$, we have ${\hat E}[f(A)\in f(\Delta)] = {\hat
E}[A\in f^{-1}(f(\Delta))]$, and hence ${\hat E}[A \in \Delta]
\leq {\hat E}[f(A)\in f(\Delta)]$. This gives as a sufficient
condition for an arrow $f_{{\cal O}}:{\hat B}\rightarrow {\hat
A}$ to be in $\nu^{\psi}(A \in \Delta)$, that $f(s(\psi,{\hat
A})) \subseteq f(\Delta)$. For suppose that $f(s(\psi,{\hat A}))
\subseteq f(\Delta)$. Then ${\hat E}[A\in s(\psi,{\hat A})] \leq
{\hat E}[f(A)\in f(s(\psi,{\hat A}))] \leq {\hat E}[f(A)\in
f(\Delta)]$. So since ${\hat E}[A \in s(\psi,{\hat A})]\psi =
\psi$, we have ${\hat E}[f(A)\in f(\Delta)]\psi = \psi$.

This condition, that $f(s(\psi,{\hat A})) \subseteq f(\Delta)$,
is also necessary. For since ${\hat E}[f(A)\in f(\Delta)] = {\hat
E}[A\in f^{-1}(f(\Delta))]$, we have that an arrow $f_{{\cal
O}}:{\hat B}\rightarrow {\hat A}$ is in $\nu^{\psi}(A \in
\Delta)$ if and only if $f^{-1}(f(\Delta)) \supseteq s(\psi,{\hat
A})$. But applying $f$ to this last we get: $f(f^{-1}(f(\Delta)))
= f(\Delta) \supseteq f(s(\psi,{\hat A}))$.

Thus we have the result (strictly in ${\cal O}_d$, and in ${\cal
O}$, for those ${\hat A}, {\hat B}, \Delta$ for which
measure-theoretic difficulties do not arise):
\begin{equation}
\nu^{\psi}(A \in \Delta) = \{f_{\cal O}:{\hat B}\rightarrow {\hat
A}: f(\Delta) \supseteq f(s(\psi,{\hat A})) \}.\label{NewCharac}
\end{equation}

This argument can be adapted to $\nu^\rho$ and $s(\rho,{\hat
A})$. We use the fact that $f_{\cal O}:{\hat B}\rightarrow {\hat
A} \in \nu^\rho(A \in \Delta)$ if and only if ${\rm tr}[\rho{\hat
E}[A \in f^{-1}(f(\Delta))]] = 1 $ if and only if
$f^{-1}(f(\Delta)) \supseteq s(\rho,{\hat A})$; which, applying
$f$, implies that $f(\Delta) \supseteq f(s(\rho,{\hat A}))$. So
we get the result (again, strictly in ${\cal O}_d$; and in ${\cal
O}$, measure-theoretic difficulties apart):
\begin{equation}
\nu^{\rho}(A \in \Delta) = \{f_{\cal O}:{\hat B}\rightarrow {\hat
A}: f(\Delta) \supseteq f(s(\rho,{\hat A})) \}.
\label{eqn:NewCharacRho}
\end{equation}

Each of Eqs.\ (\ref{NewCharac}) and (\ref{eqn:NewCharacRho}) is
clearly an analogue of part (c) of Theorem
\ref{Theorem:Val=Sieves}, which was
\begin{equation}
\alpha_{V_1}({\hat P}) = \{i_{V_2V_1}:V_2\rightarrow V_1 \mid
{\bf G}(i_{V_2V_1})(s(\alpha,V_1)) \leq {\bf G}(i_{V_2V_1})({\hat
P}) \}. \label{eqn:characterization2}
\end{equation}
and of part (c) of Theorem \ref{Theorem:alphaSubobjects},which was
\begin{equation}
\alpha_{V_1}({\hat P}) = \{i_{V_2V_1}:V_2\rightarrow V_1 \mid
{\bf I}^{\alpha}(V_1)|_{V_2} \subseteq V_1({\hat P})|_{V_2} \}
\label{eqn:characterization-interval2}
\end{equation}
In short we see that (i) $\hat A$ corresponds to $V_1$; (ii)
$\Delta$ corresponds to $\hat P$; (iii) $f$ corresponds to
coarse-graining by ${\bf G}(i_{V_2V_1})$ in Theorem
\ref{Theorem:Val=Sieves}, and by restriction to $V_2$ in Theorem
\ref{Theorem:alphaSubobjects}; and (iv) elementary supports
correspond to the rigorous supports in Theorem
\ref{Theorem:Val=Sieves} and to the intervals in Theorem
\ref{Theorem:alphaSubobjects}.

We note incidentally that the fact that $\psi \in {\cal H}$ is determined 
by the set of `certainly true' pairs $\langle {\hat A}, \Delta
\rangle$ ({\em i.e.} the pairs for which $\nu(A \in \Delta) = \,\,
\downarrow{\hat A}$), together with the fact that $\psi$ itself
determines $\nu = \nu^\psi$ by Eq.\ (\ref{eqn:nupsionO}), implies
that $\nu = \nu^\psi$ is determined by  the set of `certainly
true' pairs $\langle {\hat A}, \Delta \rangle$. This
`two-step-determination' argument (going via $\psi$) shows that
for pure quantum states $\psi$, one of the sieve-valued
valuations $\nu^\psi$ (a sieve-valued valuation that is induced
by some or other $\psi$ according to Eq.\ (\ref{eqn:nupsionO}) is
determined by the `certainly true' {\em i.e.} true$_A$
assignments that it makes.

\subsubsection{Supports give subobjects of ${\bf \Sigma}$ on ${\cal O}$}
We recall that assigning to each ${\hat A} \in {\cal O}$  a
subset $a({\hat A})$ of its spectrum $\spec A$ gives a subobject
of \SP\ (rather than the global elements prohibited by the
Kochen-Specker theorem) provided the assignment obeys the
`subset' version of $FUNC$: viz.,
\begin{equation}
 f(a({\hat A})) \subseteq a(f({\hat A})).
 \label{Defn:FUNCsset}
\end{equation}
In particular, elementary supports, as defined in Definition
\ref{Def:ElementarySupport}, induce subobjects of \SP---{\em
i.e.} interval valuations obeying Eq.\ (\ref{Defn:FUNCsset}). For
even if $\hat A$ has in part a continuous spectrum, the subset
conditions:
\begin{equation}
f(s(\psi, {\hat A})) \subseteq s(\psi, f({\hat A})); f(s(\rho,
{\hat A})) \subseteq s(\rho, f({\hat A}))\label{FuncSupportSubset}
\end{equation}
hold. So each of the interval valuations defined by
\begin{equation}
a^{\psi}({\hat A}) := s(\psi, {\hat A}); a^{\rho}({\hat A}) := s(\rho, 
{\hat A}))\label{Defn:apsi}
\end{equation}
is indeed a subobject of \SP. If $\hat A$ has pure discrete
spectrum, Eq.\ (\ref{FuncSupportSubset}) becomes an equality, both for a 
vector state and a density matrix:
\begin{equation}
f(s(\psi, {\hat A})) = s(\psi, f({\hat A})); f(s(\rho, {\hat A}))
= s(\rho, f({\hat A})).\label{FuncSupportEqual}
\end{equation}

\section{Defining sieve-valued valuations in terms of subobjects of $\bf
\Sigma$} \label{Sec:ValSubobjects} As we have seen, the
correspondence, indeed mutual determination, in Section
\ref{Sec:interval} between sieve-valued and interval-valued
valuations holds for a wider class of valuations than just those
discussed in Sections \ref{ssec:sievegenval} to
\ref{SubSec:IntervalValuations}. In particular, Theorems
\ref{Theorem:Val=Sieves} and \ref{Theorem:alphaSubobjects} used
only the first clause of the definition in Section
\ref{ssec:sievegenval} of a sieve-valued valuation (Definition
\ref{Defn:gen-val-V}), viz.\ the requirement that a sieve-valued
valuation obey $FUNC$. This situation suggests that it would be
worth surveying different ways of defining sieve-valued and
interval-valued valuations---and the properties that ensue from
these definitions. In this Section we undertake a part of such a
survey. It will show in particular that the valuations we have
considered are a very natural way to secure the properties listed
in the other clauses of Definition \ref{Defn:gen-val-V}.

To be precise: we will focus on the role played in the results of
Section \ref{Sec:interval} by our having defined generalised
valuations (with sets of morphisms as values) in terms of the
partial order relation at each stage. That is, we note that in
the discussion of global elements of \G\ in Section
\ref{sssec:globalG}, the results about such valuations repeatedly
invoked the partial order $\leq$ in $\G(V):= {\cal L}(V)$ (cf.\
in particular, (i) of Theorem \ref{Theorem:Val=Sieves}); and in
the discussion in Section \ref{SubSubSec:SubObjectSigma} of
subobjects of \SP, the results repeatedly invoked subsethood
$\subseteq$ among subsets of the spectrum $\sigma(V)$ (cf.\ (i)
of Theorem \ref{Theorem:alphaSubobjects}). In both cases, the
relation is used at $V_2$, the coarser of two stages $V_2
\subseteq V_1$, to connect a notion `intrinsic' to $V_2$ ({\em
i.e.}, $s(\alpha, V_2))$ and ${\bf I}^\alpha(V_2)$ respectively)
to a notion got by coarse-graining from $V_1$ ({\em i.e.},
$\G(i_{V_2V_1})({\hat P})$ and $V_1({\hat P})|_{V_2} =
V_2(\G(i_{V_2V_1})({\hat P}))$ respectively).

So we will now ask how the properties of valuations taking sets
of morphisms as values that are defined in terms of interval
valuations by using a relation $R$ to connect a notion intrinsic
to a stage $V_2$ to another notion got by coarse-graining from a
finer stage $V_1$, depend upon the choice of the relation $R$.
That is to say, we will now consider the following schema for
defining from a given global element $a$ of $\G$, a valuation
taking sets of morphisms as values, in terms of an arbitrary
binary relation $R$:
\begin{equation}
\alpha^{a,R}_{V_1}({\hat P}):= \{i_{V_2V_1}: V_2 \rightarrow V_1
\mid a(V_2)\; R \; {\bf G}(i_{V_2V_1})({\hat P})\}.
\label{eqn:intervaltosieveR}
\end{equation}
The analogous general schema starting from an interval valuation
$a$ that is a subobject of \SP\ is
\begin{equation}
\alpha^{a,R}_{V_1}({\hat P}):= \{i_{V_2V_1}: V_2 \rightarrow V_1
\mid a(V_2) \; R \; V_1({\hat P})\mid_{V_2} \}.
\label{eqn:intervalassubobjectSigmatosieveR}
\end{equation}
Similarly, for the case of $\cal O$ (cf.\ Section
\ref{SubSec:CorrespondenceO}), the general schema is
\begin{equation}
\alpha^{a,R}_{\hat A}({\Delta}):= \alpha^{a,R}(A \in \Delta) :=
\{f_{\cal O}: {\hat B} \rightarrow {\hat A} \mid a({\hat B}) \; R
\; f( \Delta) \}. \label{eqn:intervalonOtosieveR}
\end{equation}
But we will discuss only Eq.\ (\ref{eqn:intervaltosieveR}); our
results carry over {\em mutatis mutandis} to the cases of Eqs.\
(\ref{eqn:intervalassubobjectSigmatosieveR}) and
(\ref{eqn:intervalonOtosieveR}).

This leads us to ask what conditions on $R$ in Eq.\
(\ref{eqn:intervaltosieveR}) correspond, as either necessary or
sufficient conditions, to various properties of the valuation
$\alpha^{a,R}$? The following results can be immediately
verified. We give them in the same order as the conditions listed
in our original definition of a sieve-valued valuation (Defn.\
\ref{Defn:gen-val-V}. in Section \ref{ssec:sievegenval}).

(i) $\alpha^{a,R}_{V_1}({\hat P})$ is a sieve if and only if $R$
is {\em stable under coarse-graining\/} in the sense that if $
a(V_2)\; R\; \G(i_{V_2V_1})({\hat P})$, then for all $V_3
\subseteq V_2,\, a(V_3)\; R\; \G(i_{V_3V_1})({\hat P})$.

Since $a$ is assumed to be a global element of \G, so that for
all $V' \subseteq V$, $a(V') = \G(i_{V'V})a(V)$, the consequent
in this condition becomes
\begin{equation}
a(V_3) = \G(i_{V_3V_2})a(V_2) \; R\; \G(i_{V_3V_1})({\hat P}) =
\G(i_{V_3V_2})\G(i_{V_2V_1})({\hat P}).
\end{equation}
Since \G\ is monotonic with respect to $\leq$, choosing $R$ to be
$\leq$, as we have done (cf.\ Eqs.\
(\ref{eqn:alphasupportsmatch}) and (\ref{eqn:characterization}))
is a very natural way to secure sievehood.

(ii) For any relation $R$ whatsoever, $\alpha^{a,R}$ obeys
functional composition in the form of Eq.\ (\ref{eqn:FC-alpha}),
{\em i.e.}
\begin{equation}
\alpha^{a,R}_{V_2}({\bf G}(i_{V_2V_1}(\hat P))=
i_{V_2V_1}^*(\alpha^{a,R}_{V_1}(\hat P)). \label{eqn:FC-alphaR}
\end{equation}
To see this, note that in the argument of part (b) of Theorem
\ref{Theorem:Val=Sieves}, any relation $R$ could be substituted
for $\leq$.

(iii) $\alpha^{a,R}$ obeys the null proposition condition, {\em
i.e.} $\alpha^{a,R}_{V_1}({\hat 0}) = \emptyset$, if and only if
there is no $V_2 \subseteq V_1$ with $a(V_2) \, R \,\, {\hat 0}$
(since $\hat 0$ coarse-grains to $\hat 0$). Provided $a$ always
assigns a non-zero projector, this condition is satisfied by our
choice of $R$ as $ \leq$.

(iv) $\alpha^{a,R}$ obeys the monotonicity condition, i.e. if
${\hat P} \leq {\hat Q} \in {\cal L}(V_1)$ then
$\alpha^{a,R}_{V_1}({\hat P}) \leq \alpha^{a,R}_{V_1}({\hat Q})$,
if and only if $R$ is {\em isotone under coarse-graining\/} in
the sense that
\begin{equation}
[{\hat P} \leq {\hat Q}  \mbox{ and }
a(V_2)\,R\,\G(i_{V_2V_1})({\hat P})] \Rightarrow
a(V_2)\,R\,\G(i_{V_2V_1})({\hat Q}).
\end{equation}
Since $\G$ is monotonic with respect to $\leq$, the natural
sufficient condition for this is that the relation $R$ is stable
under taking larger elements on its left-hand side, i.e.
\begin{equation}
[{\hat S} \leq {\hat T} \mbox{ and } a(V_2) \, R \, {\hat S}]
\Rightarrow a(V_2) \, R \, {\hat T}.
\end{equation}
Again, the natural choice for satisfying this is that $R$ is
taken to be $\leq$.

(v) $\alpha^{a,R}$ obeys the exclusivity condition, that if
${\hat P}{\hat Q}= 0$ and $\alpha^{a,R}_{V_1}({\hat P}) =
\downarrow_{V_1} = \mbox{ true}_{V_1}$, then
$\alpha^{a,R}_{V_1}({\hat Q}) < \mbox{ true}_{V_1}$, if and only
if
\begin{eqnarray}
\lefteqn{ \mbox{If }{\hat P}{\hat Q}= 0 \makebox{ and }\forall
V_2 \subseteq V_1, a(V_2)\,R\,\G(i_{V_2V_1})({\hat P}), \makebox{
then } \exists V_3
\subseteq V_1}\hspace{5cm} \nonumber\\
&&\makebox{ such that not } a(V_3)\,R\,\G(i_{V_3V_1})({\hat Q}).
\end{eqnarray}
 Here, the
condition in terms of $R$ is not very different from exclusivity
in the original form; and so seems not very illuminating.  But
provided $a$ always assigns a non-zero projector, this condition
is satisfied by our choice of $R$ as $ \leq$. For if ${\hat
P}{\hat Q}= 0$ and, for all $V_2 \subseteq V_1$ we have
$a(V_2)\leq \,\G(i_{V_2V_1})({\hat P})$, then $a(V_1) \leq {\hat
P}$, so that $\neg a(V_1) \leq {\hat Q}$ (since $a(V_1) \neq
{\hat 0}$), and hence $\alpha^{a,{\leq}}_{V_1}({\hat Q}) \neq
\mbox{ true}_{V_1}$.

(vi) $\alpha^{a,R}$ obeys the unit proposition condition, that
$\alpha^{a,R}_{V_1}({\hat 1})= \mbox{ true}_{V_1}$, if and only
if for all $V_2 \subseteq V_1$ we have
$a(V_2)\,R\,\G(i_{V_2V_1})({\hat 1})= {\hat 1}_{V_2}$ (since
${\hat 1}$ coarse-grains to ${\hat 1}$). Again, the natural way
for satisfying this is to choose the relation $R$ to be $\leq$.

These results show that there is a natural choice of the relation
$R$, viz.\ $R:= \leq$, which is sufficient to yield {\em all\/}
of the properties (i.e. clauses  (i)--(v) of Definition
\ref{Defn:gen-val-V}), provided $a$ always assigns a non-zero
projector. And again this conclusion reflects the theme of
Section \ref{Sec:interval}, viz.\ the correspondence between
sieve-valued and interval-valued valuations, and in particular
the characterization Eq.\ (\ref{eqn:characterization}) (part (c)
of Theorem \ref{Theorem:Val=Sieves}).

Furthermore, analogous results are easily verified for the
schemas in Eqs.\ (\ref{eqn:intervalassubobjectSigmatosieveR}) and
(\ref{eqn:intervalonOtosieveR}). More precisely: taking the
relation $R$ as subsethood, $\subseteq$ in these schemas is
sufficient for these properties, provided some `regularity
conditions' hold. These conditions include:
\begin{enumerate}
\item The analogue of the proviso above, that $a$ always assigns a 
non-zero projector ({\em i.e.} that $a$ always assigns a non-empty
subset).

\item The requirement that $a$ defines a `tight' subobject of \SP, in the
sense of Eq.\ (\ref{eqn:intervaltight}).

\item For the case of $\cal O$, ({\em i.e.} schema
\ref{eqn:intervalonOtosieveR}) for all bounded Borel functions
$f$ and all $\hat A$, we have the equality $f(\sigma(\hat A)) =
\sigma(f(\hat A))$ (as always occurs if $\hat A$ has pure
discrete spectrum), not merely $f(\sigma(\hat A)) \subseteq
\sigma(f(\hat A))$ as in Eq.\ (\ref{eqn:Ospectraonlysubset}).
\end{enumerate}
But we will not go into details of just how these regularity
conditions make choosing $R$ as subsethood $\subseteq$ sufficient
for the various properties (i)--(vi) above, for the schemas of
Eqs.\ (\ref{eqn:intervalassubobjectSigmatosieveR}) and
(\ref{eqn:intervalonOtosieveR}). But again the conclusion---that
taking $R$ as subsethood in these schemas is sufficient for these
properties---reflects the correspondence in Section
\ref{Sec:interval} between sieve-valued and interval-valued
valuations; and in particular the characterizations, Eq.\
(\ref{eqn:characterization-interval}) (for $\cal V$: part (c) of
Theorem \ref{Theorem:alphaSubobjects}), and Eqs.\
(\ref{NewCharac}) and (\ref{eqn:NewCharacRho}) (for $\cal O$).

\section{Conclusion}
In this paper, we have extended our topos-theoretic perspective
on the assignment of values to quantities in quantum theory;
principally using the base category $\cal V$ of commutative von
Neumann algebras introduced in \cite{HIB00}. In Section
\ref{Sec:interval}, we compared our sieve-valued valuations with
interval valuations based on the notion of supports. This
discussion (adding to some results reported in Section 4 of
\cite{HIB00}) had as its main theme a correspondence (mutual
determination) between certain sieve-valued valuations and
corresponding interval valuations. This correspondence was summed
up (for $\cal V$) in the characterizations given in parts (c) of
Theorems \ref{Theorem:Val=Sieves} and
\ref{Theorem:alphaSubobjects}, Eqs.\ (\ref{eqn:characterization})
and (\ref{eqn:characterization-interval}); and summed up more
heuristically for $\cal O$, in Eqs.\ (\ref{NewCharac}) and
(\ref{eqn:NewCharacRho}).

In Section \ref{Sec:ValSubobjects}, we generalized this
discussion: we gave a partial survey of how in defining
sieve-valued valuations in terms of interval valuations, certain
properties of the sieve-valued valuations derive from the
properties of the binary relation $R$ used in the definition.
This survey again showed the naturalness of our previous
definitions. For taking $R$ to be the partial order $\leq$ among
projectors, or to be subsethood $\subseteq$ among subsets of
spectra, was a natural and simple sufficient condition for the
defined valuations to obey the clauses of our original definition
of sieve-valued valuations (Definition \ref{Defn:gen-val-V}).
\\
\\

\noindent {\bf Acknowledgements} Chris Isham gratefully
acknowledges support by the EPSRC grant GR/R36572. Jeremy Butterfield 
thanks Hans Halvorson for very helpful discussions.


\begin{thebibliography}{10}

\bibitem{B01}
J.~Butterfield.
\newblock Topos Theory as a Framework for Partial Truth.
\newblock ``Proceedings of the 11th International Congress of Logic
Methodology and Philosophy of Science'', ed. P. Gardenfors, K.
Kijania-Placek and J. Wolenski , pp. ????, Kluwer Academic, 2001.
Available at: http://philsci-archive.pitt.edu, ID code =
PITT-PHIL-SCI00000192.


\bibitem{IB98}
C.J.~Isham and J.~Butterfield.
\newblock A topos perspective on the {K}ochen-{S}pecker theorem:
{I.} {Q}uantum states as generalised valuations.
\newblock {\em Int.\ J.\ Theor.\ Phys.}, {\bf 37},
2669--2733, 1998.


\bibitem{IB99}
J.~Butterfield and C.J.~Isham.
\newblock A topos perspective on the {K}ochen-{S}pecker theorem:
{II.} {C}onceptual aspects, and classical analogues.
\newblock  {\em Int.\ J.\ Theor.\ Phys.}, {\bf 38}, 827--859, 1999.

\bibitem{HIB00}
J.~Hamilton, J.~Butterfield and C.J.~Isham.
\newblock A topos perspective on the {K}ochen-{S}pecker theorem:
{III.} {V}on {N}eumann algebras as the base category.
\newblock  {\em Int.\ J.\ Theor.\ Phys.}, {\bf 39}, 1413-1436, 2000.


\bibitem{KS67}
S.~Kochen and E.P.~Specker.
\newblock The problem of hidden variables in quantum mechanics.
\newblock {\em Journal of Mathematics and Mechanics}, {\bf17} 59--87,
1967.

\bibitem{KR83a}
R.V.~Kadison and J.R.~Ringrose.
\newblock {\em Fundamentals of the Theory of Operator Algebras
Volume 1: Elementary Theory}
\newblock Academic Press, New York, 1983.

\bibitem{Verm00}
P.~Vermaas
\newblock {\em A Philosopher's Understanding of Quantum Mechanics}
\newblock Cambridge University Press, Cambridge UK, 2000.

\end{thebibliography}
\end{document}